\documentclass{aa}

\usepackage{graphicx}
\usepackage{natbib}

\usepackage{txfonts}

\begin{document}

   \title{Distance and extinction to the Milky Way spiral arms along the Galactic centre line of sight.}
   \titlerunning{Distance and extinction to the Milky Way spiral arms along the Galactic centre line of sight.}
   \authorrunning{Nogueras-Lara et al.}

  \author{F. Nogueras-Lara
          \inst{1}
          \and      
          R. Sch\"odel 
          \inst{2}       
          \and
          N. Neumayer 
          \inst{1}    
          }

   \institute{
    Max-Planck Institute for Astronomy, K\"onigstuhl 17, 69117 Heidelberg, Germany
              \email{nogueras@mpia.de}
       \and 
           Instituto de Astrof\'isica de Andaluc\'ia (IAA-CSIC),
     Glorieta de la Astronom\'ia s/n, 18008 Granada, Spain                         
       }
   \date{}

  \abstract
   {The position of the Sun inside the disc of the Milky Way significantly hampers the study of the spiral arm structure given the high amount of dust and gas along the line of sight, and the overall structure of this disc has therefore not yet been fully characterised.}
   {We aim to analyse the spiral arms in the line of sight towards the Galactic centre (GC) in order to determine their distance, extinction, and stellar population.}
   {We use the GALACTICNUCLEUS survey, a $JHK_s$ high-angular-resolution photometric catalogue ($0.2''$) for the innermost regions of the Galaxy. We fitted simple synthetic colour-magnitude models to our data via $\chi^2$ minimisation. We computed the distance and extinction to the detected spiral arms. We also analysed the extinction curve and the relative extinction between the detected features. Finally, we studied extinction-corrected $K_s$ luminosity functions (KLFs) to study the stellar populations present in the second and third spiral arm features.
}
{We determined the mean distances to the spiral arms: $d_1 = 1.6\pm 0.2$ kpc, $d_2 = 2.6\pm 0.2$ kpc, $d_3 = 3.9\pm0.3$ kpc, and $d_4 = 4.5 \pm 0.2$ kpc, and the mean extinctions: $A_{H1} = 0.35\pm0.08$ mag, $A_{H2} = 0.77\pm0.08$ mag, $A_{H3} = 1.68\pm0.08$ mag, and $A_{H4} = 2.30\pm0.08$ mag. We analysed the extinction curve in the near-infrared for the stars in the spiral arms and find mean values of $A_J/A_{H} = 1.89 \pm 0.11$ and $A_H/A_{K_s} = 1.86 \pm 0.11$, in agreement with the results obtained for the GC. This implies that the shape of the extinction curve does not depend on distance or absolute extinction. We also built extinction maps for each spiral arm and find them to be homogeneous and that they might correspond to independent extinction layers. Finally, analysing the KLFs from the second and third spiral arms, we find that they have similar stellar populations. We obtain two main episodes of star formation: $>6$\,Gyr ($\sim 60-70\,\%$ of the stellar mass), and $1.5-4$\,Gyr ($\sim 20-30\,\%$ of the stellar mass), compatible with previous work. We also detect recent star formation at a lower level ($\sim10\%$) for the third spiral arm.
}
   
   {}


\titlerunning{Distance and extinction to the Milky Way spiral arms along the Galactic centre line of sight}
\authorrunning{F. Nogueras-Lara et al.}

   \maketitle
%

 \section{Introduction}
 \label{intro_sect}
 
The position of the Sun in the Galactic disc allows us to perform a detailed analysis of the stellar population in its close vicinity \citep[e.g.][]{Bland-Hawthorn:2016aa,Ruiz-Lara:2020aa}. Nevertheless, this position significantly hampers the study of the spiral arm structure of the  Milky Way and its properties,  given the interstellar dust that characterises the low-Galactic-latitude lines of sight \citep[e.g.][]{Chen:2017aa}. However, the characterisation of the spiral arms of the Milky Way is of fundamental interest because it is needed in order to understand our Galaxy in a wider context of galactic morphology, dynamics, and evolution.

Much effort during recent decades has gone into properly characterising the main parameters (number, shape, inter-arm separation) of the spiral arm structure \citep[e.g.][and references therein]{Bland-Hawthorn:2016aa}. Nevertheless, the whole structure remains unclear \citep[e.g.][]{Momany:2006aa,Hou:2014aa,Khoperskov:2020aa}. According to recent studies based on trigonometric parallaxes and proper motions of molecular masers associated with young high-mass stars, the Milky Way appears as a four-arm spiral galaxy with some extra segments and spurs \citep{Reid:2019aa}.

In particular, the analysis of the line of sight towards the Galactic centre (GC) is hampered by the extreme cumulative extinction \citep[with an average value of $A_V\gtrsim30$ mag, corresponding to $A_{K_s}\gtrsim2.5$ mag, see e.g.][]{Nishiyama:2008qa,Schodel:2010fk,Fritz:2011fk,Nogueras-Lara:2018aa,Nogueras-Lara:2020aa} due to the dust and gas present in the Galactic disc. Therefore, the analysis of its stellar population is restricted to infrared wavelengths, which suffer least from interstellar reddening. The precise measurements of the Gaia survey \citep[e.g.][]{Gaia-Collaboration:2018aa} within the Galactic disc are limited to distances $\lesssim 3$\,kpc from the Sun, therefore most studies of the Milky Way disc structure used radio observations of masers and/or star forming regions; their distances are measured by means of trigonometric parallaxes \citep[e.g.][]{Reid:2016aa,Bland-Hawthorn:2016aa,Wu:2019aa, Reid:2019aa}. Moreover, infrared surveys such as the COBE/DIRBE survey and the infrared Spitzer/GLIMPSE survey have been used to trace the spiral arm structure \citep[e.g.][]{Drimmel:2001aa,Churchwell:2009fk}.

The analysis of red clump (RC) stars \citep[red giant stars in their helium-core-burning sequence, e.g. ][]{Girardi:2016fk} using near infrared (NIR) photometry is key to analysing the structure and stellar population towards the innermost regions of the galaxy. In this way, NIR data from the 2MASS \citep{Skrutskie:2006te} and UKIDSS \citep{Lucas:2008ut} surveys have been used to trace the spiral structure and to outline the shape of the Galactic bulge and bar \citep[e.g.][]{Cabrera-Lavers:2008kx,Francis:2012vt,Robin:2012wn}. More recently, the VVV survey \citep{Minniti:2010fk,Saito:2012ml} allowed improvement of the study towards the obscured regions of the Galactic bar \citep[e.g.][]{Gonzalez:2011aa,Minniti:2014aa}, and also permitted the spiral arm structure to be traced beyond the Galactic bulge \citep[e.g.][]{Gonzalez:2018aa,Saito:2020vp}.

With the present study, we aim to characterise the spiral arms along the line of sight towards the GC, computing their distance and the extinction using NIR photometry from the GALACTICNUCLEUS survey \citep{Nogueras-Lara:2018aa,Nogueras-Lara:2019aa}. It is specially designed to characterise the structure and the stellar population of the nuclear bulge of the Milky Way with unprecedented detail. The high angular resolution of the data and the fact that it is several magnitudes deeper than any existing NIR catalogue for the GC \citep[e.g.][]{Nogueras-Lara:2018aa,Nogueras-Lara:2019aa} make this survey the best of its kind and the only one that reveals the details of the structure of the  Galactic disc towards this extremely crowded and highly extinguished line of sight. We also used the Gaia survey to cross-check our results for the stars belonging to the closest spiral arms.

 \section{Data}

For the analysis presented in this paper, we used the GALACTICNUCLEUS catalogue \citep{Nogueras-Lara:2018aa,Nogueras-Lara:2019aa}. This is a NIR $JHK_s$ survey carried out with the HAWK-I instrument \citep{Kissler-Patig:2008fr} at the ESO VLT unit telescope\,4. It is a single-epoch catalogue consisting of 49 pointings in all three bands ($JHK_s$), and includes observations of more than three million sources in the  nuclear stellar disc (NSD), the innermost Galactic bulge, and the transition region between the bulge and the NSD. It uses the speckle holography technique \citep{Schodel:2013fk} to achieve high angular resolution of $\sim 0.2''$, reaching 5\,$\sigma$ detections at $J\sim22$ mag, $H\sim21$\,mag, and $K_s\sim 21$\,mag. The survey supersedes all previous photometric surveys for the GC region by several magnitudes. The uncertainties on the photometry are below 0.05\,mag at $J\sim21$\,mag, $H\sim19$\,mag, and $K_s\sim 18$\,mag. The zero point (ZP) systematic uncertainty is below 0.04\,mag in all three bands.

\begin{figure}
   \includegraphics[width=\linewidth]{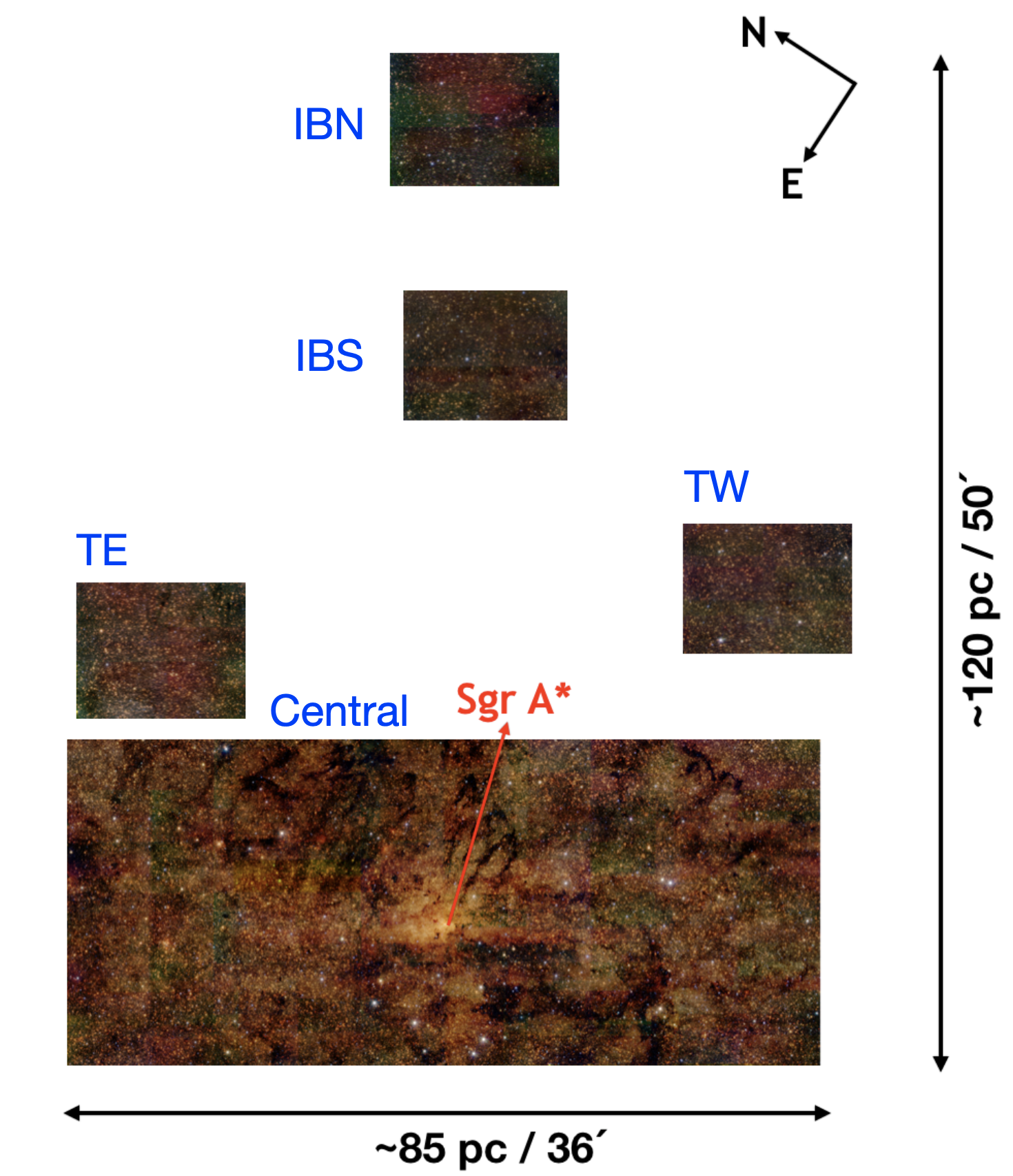}
   \caption{Red giant branch image from the GALACTICNUCLEUS survey (using $JHK_s$ bands) showing the regions used in this study. The red arrow marks the position of the supermassive black hole, Sagittarius A* (Sgr\,A*). The names of the regions are also indicated in the figure: central, transition east (TE), transition west (TW), inner bulge south (IBS), and inner bulge north (IBN). The distances between the regions correspond to real distances, as indicated by the scaling arrows.}

   \label{GNS}
    \end{figure}

As  we aim to analyse the stellar population between the Earth and the GC, we used five regions from GALACTICNUCLEUS that correspond to different lines of sight given their different Galactic latitudes. In particular, we used the central region, the transition regions east (TE) and west (TW), and the inner bulge regions north (IBN) and south (IBS). Figure\,\ref{GNS} shows a scheme of the studied area. Given the relatively low number of stars in the TE, TW, IBS, and IBN regions, we combined the transition regions and the inner bulge ones to end up with central, transition, and inner bulge regions.

\begin{figure}
   \includegraphics[width=\linewidth]{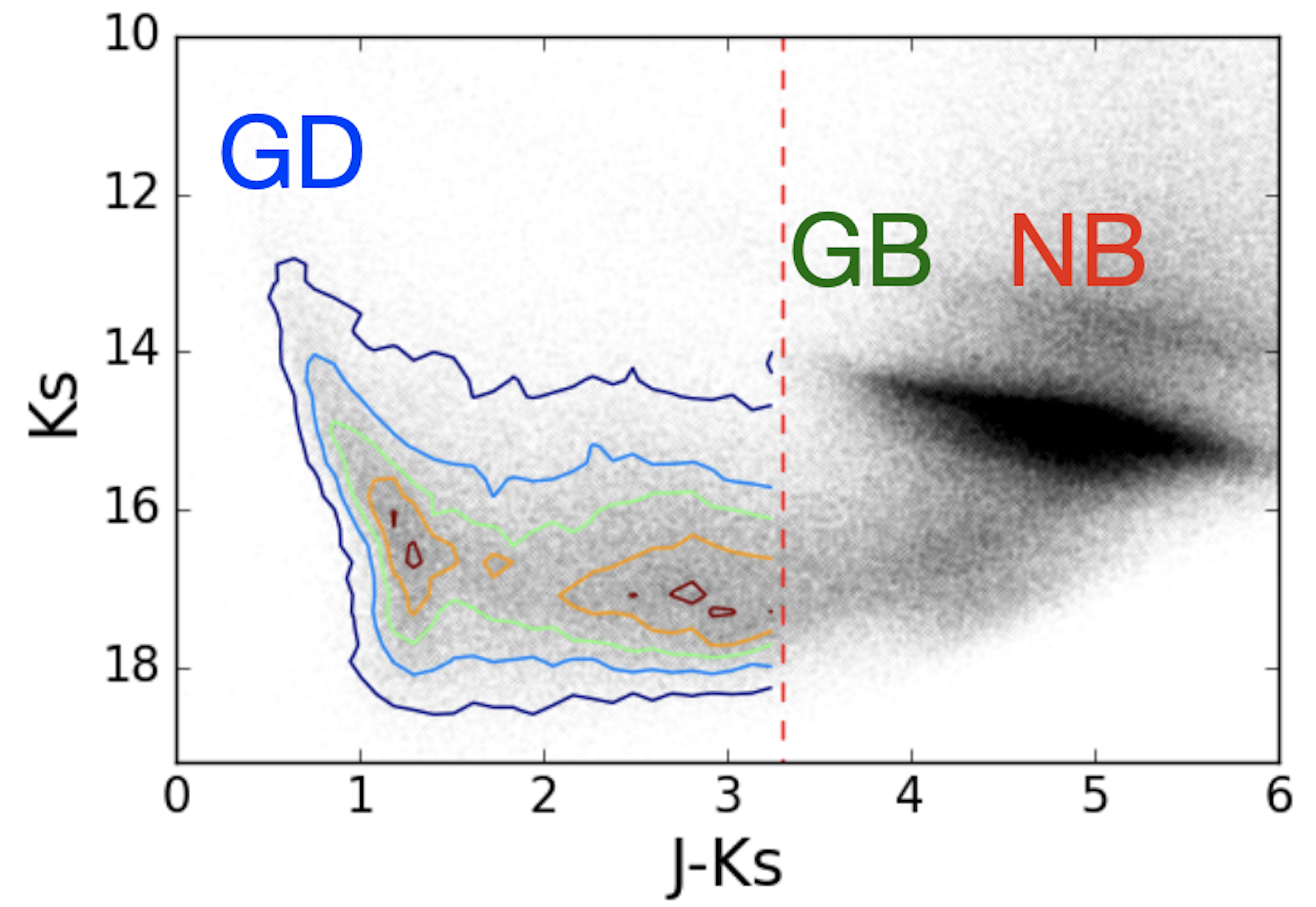}
   \caption{Colour-magnitude
diagram $K_s$ vs. $J-K_s$ from the central region of the GALACTICNUCLEUS survey. The red dashed line indicates the colour cut between the stellar population belonging to the GC: Galactic bulge (GB) + Nuclear bulge (NB), and the Galactic disc (GD). The contours outline the population belonging to the GD.}

       \label{CMD}
    \end{figure}

The stellar population belonging to the GC is highly reddened \citep[e.g.][]{Schodel:2010fk,Nogueras-Lara:2018aa, Nogueras-Lara:2019aa}, which allows us to clearly distinguish it from the foreground population belonging to the stellar disc using a simple colour cut \citep{Nogueras-Lara:2019ad}. Figure \ref{CMD} shows the colour-magnitude diagram (CMD) $K_s$ versus $J-K_s$ of the central region of the GALACTICNUCLEUS survey. This region allows us to easily differentiate between the stellar disc and the GC because it is the most extinguished region and the one with the largest number of stars in the GALACTICNUCLEUS catalogue. The most prominent feature (around $K_s\sim$15, $J-K_s\sim5$ mag in the upper panel) corresponds to RC stars belonging to the GC \citep[see Fig. 14 in][]{Nogueras-Lara:2019aa}. Namely, the GC stellar population comprises stars from the nuclear bulge \citep[NB, e.g.][]{Launhardt:2002nx,Nogueras-Lara:2019ad} and the inner Galactic bulge \citep[e.g.][]{Nogueras-Lara:2018ab}, the stellar population of this latter being less reddened than the one pertaining to the NB (see Fig. \ref{CMD} for details). Stars located at redder colours correspond to the Galactic disc. In this way, and as an initial assumption that is later corroborated by our analysis, the red dashed line in the figure indicates the approximate colour division between the foreground population and GC stars.



\section{Colour-magnitude
diagram fitting}
\label{cmd_fitting}

Previous work using the GALACTICNUCLEUS data suggested the presence of spiral arms towards the GC by visual inspection of the CMDs \citep[e.g. Figs. 14 and 14 from][respectively]{Nogueras-Lara:2018aa,Nogueras-Lara:2019aa}. A number of studies based on different spiral arm tracers detect four spiral arms along the line of sight towards the GC \citep[e.g.][]{Hou:2014aa,Vallee:2016aa,Reid:2019aa}. In this way, we analysed our data, constructing a simple synthetic model considering the four spiral arms scenario as a fiducial model.

\subsection{Models}
\label{mod}

We created synthetic models for each of the spiral arms assuming the star formation history (SFH) derived by \citet{Ruiz-Lara:2020aa} for the thin disc in a 2-kpc-radius bubble around the Sun. In this way, we combined stellar populations of 13, 6, 2, 1, and 0.1\,Gyr and scaled them in agreement with the SFH shown in Fig. 4 of \citet{Ruiz-Lara:2020aa}. The metallicity of the models was selected taking into account the age--metallicity relation for the solar neighbourhood, assuming the mean values for different age distributions from Fig. 18 in \citet{Feltzing:2001aa}. In particular, we used $[M/H]=-0.07$ for all the models except for the oldest one ($[M/H]=-0.28$). This metallicity distribution is also in agreement with previous studies of the Galactic disc \citep{Bergemann:2014aa,Ruiz-Lara:2020aa}. 

We used PARSEC models \citep[release v1.2S + COLIBRI S\_35, ][]{Bressan:2012aa,Chen:2014aa,Chen:2015aa,Tang:2014rm,Marigo:2017aa,Pastorelli:2019aa} and created the synthetic CMDs using the web tool CMD\,3.3\footnote{http://stev.oapd.inaf.it/cgi-bin/cmd}. We used the CMD $H$ versus $J-H$ for the model fitting because in this way we avoid the $K_s$ band that is more affected by saturation and crowding than the $H$ band. Additionally, using the CMD $H$ versus $J-H$ we can more clearly separate the different components along the line of sight because their mean colour differences due to extinction are significantly larger at $J-H$ than the scatter due to measurement uncertainties \citep{Nogueras-Lara:2019aa}. The model was generated as follows:

The main parameters of the model are the distance to each spiral arm and its associated extinction ($d$ and $A_H$, respectively). The distance only affects the $y$-axis of the CMD (the colours are not distance dependent). On the other hand, to compute the extinction in the $J$ band ($A_J$) given $A_H$, we used the relation $A_H/A_J=1.87\pm0.03$ obtained by \citet{Nogueras-Lara:2020aa}.\\


We simulated the photometric uncertainties of the data assuming a Gaussian distribution with a standard deviation of 0.05 mag for each of the bands. This value considers the statistical uncertainties and also the possible ZP variations ($\sim$0.04\,mag) between the different pointings that were combined to get the final photometric catalogue \citep{Nogueras-Lara:2019aa}.\\                                            
                                                
We simulated a background of stars, whose relative contribution is a free parameter of the fit. This background follows an exponential distribution to account for the Galactic disc component. We used a scale length of 2600\,pc \citep{Bland-Hawthorn:2016aa} and placed stars at distances between 100 and 5500\,pc from the Earth. The interstellar extinction was assumed to linearly increase from $A_H = 0$ (corresponding to 0\,pc) to $A_H = 2.025$\,mag. This value was computed using the GC extinction of $A_{1.61\ \mu m} = 3.40$\,mag (at $\sim8$ kpc) obtained in \citet{Nogueras-Lara:2020aa}, and assuming that it increases linearly with the distance from the Earth to the GC.  Given the complexity of the extinction towards the innermost regions of the Galaxy \citep[e.g.][]{Nataf:2016vc,Alonso-Garcia:2017aa,Nogueras-Lara:2020aa}, we used the three-dimensional extinction maps obtained by \citet{Schultheis:2014wm} to assess the assumption of a linear increase in extinction. These authors obtained a steep rise in $A_{K_s}$ for different lines of sight near the GC, with a flattening occurring at $\sim4-6$\,kpc (see their Fig.\,8). The steep behaviour is compatible with the linear increase that we assumed, and it is probably due to the presence of gas and dust in the Galactic disc. On the other hand, the flattening is likely produced by the lower dust and gas content associated to the Galactic bulge in comparison to the Galactic disc \citep[e.g.][]{Englmaier:1999fk}. The distance $\sim4-6$\,kpc was obtained using the extinction curve obtained by \citet{Nishiyama:2009oj} ($A_J/A_{K_s}\sim3.02$). Using the more recent extinction curve by \citet{Nogueras-Lara:2020aa} ($A_J/A_{K_s}\sim3.44$) would increase the estimated distance where the flattening occurs by $\sim1$\,kpc, matching our assumption of an upper limit for the contribution of background stars at 5500\,kpc from the Earth.\\

The numbers of stars in each spiral arm were defined as a ratio to the number in the first arm.\\

We considered that the observed area increases as the square of the distance, implying that the number of stars is proportional to the square radius.\\

We assumed that all the spiral arms have the same width \citep[$\sim$400\,pc from mid-arm to dust lane,][]{Vallee:2014aa}. To simulate this, we used a Gaussian with a standard deviation of 400 pc to randomly generate the width effect for each spiral arm.\\

The extinction to each spiral arm was a free parameter and depends on each of the models considered.\\

After having generated the models, we applied a Gaussian uncertainty distribution to the simulated $J$ and $H$ photometry. This accounts for the possible differential extinction towards the features and also for some possible additional measurement uncertainties. For this, we used Gaussians with a standard deviation of 0.05\,mag.

\subsection{Fitting procedure}

For the model fit, we restricted the data to stars that belong to the foreground population as indicated in Fig. \ref{CMD}. We only used the central region, because the lower extinction towards bulge and transition regions makes it difficult to separate Galactic disc stars from GC stars and because the lower number of stars towards these regions would lead to lower quality fits. Interstellar extinction towards the GC is significantly higher at low latitudes, which facilitates the separation of disc and GC stars \citep{Schodel:2014fk,Nogueras-Lara:2018ab}.

We only considered stars with $J-H\in[0.3, 2.3]$ and $H\in[10,18.3]$ mag, where the completeness in $H$ band is more than 80\,\% according to \citet{Nogueras-Lara:2020aa}. To estimate the uncertainty of the fits and to significantly decrease the necessary computing time, we implemented a Monte Carlo (MC) simulation creating 500 independent data sets, randomly sampling 5000 of the accepted stars for each of them. 

To compare the model with the data, we binned the selected region of the CMD. We used the Python function numpy.histogram\_bin\_edges \citep{Harris:2020aa} to compute the bin widths for each axis. We ended up with 0.09 and 0.18 mag for the $J-H$ and $H$ axes, respectively.

We generated a grid of models to fit the MC samples, varying the free parameters (see Table \ref{model_table}) and identifying the best ones via $\chi^2$ minimisation. The distance parameters were selected to homogeneously sample the parameter space in steps of 350 pc. The extinction values also cover the space homogeneously in steps of 0.15 mag up to an upper limit of 2.3 mag.

We used the Poisson maximum likelihood parameter $\chi^2 = 2\Sigma_i m_i - n_i + n_i\ ln(n_i/m_i)$, where $m_i$ is the number of the stars of the model, $n_i$ is the number of stars in the observations, and the subindex $i$ indicates the bin of the CMD \citep[e.g.][]{Mighell:1999aa,Dolphin:2002aa,Pfuhl:2011uq}. Our grid approach allowed us to:  (1) not have to choose any starting parameters for spiral arm distances and extinctions, (2) explore the whole parameter space avoiding false minima that might appear when using non-linear least-squares approaches, and (3) reduce the computing time to make the computation feasible.

\begin{table}
\caption{Parameters to create the models.}
\label{model_table} 
\begin{center}
\def\arraystretch{1.4}
\setlength{\tabcolsep}{3.8pt}
\begin{tabular}{cc}
\hline 
\hline 
Parameters & \tabularnewline
\hline 
scaling factor & $S_{back}$ = 25, 70, 115, 150\%\tabularnewline
$d_1$ =1300, 1650, 2000 & $A_{H_1}$ = 0.05, 0.2, 0.35, 0.5\tabularnewline
$d_2$ = 2350, 2700, 3050 & $A_{H_2}$ = 0.65, 0.8, 0.95, 1.1\tabularnewline
$d_3$ = 3400, 3750, 4100 & $A_{H_3}$ = 1.25, 1.4, 1.55, 1.7\tabularnewline
$d_4$ = 4450, 4800, 5150 & $A_{H_4}$ = 1.85, 2.0, 2.15, 2.3\tabularnewline
\hline 
\end{tabular}

\end{center}

\textbf{Notes.} The scaling factor scales the number of stars of the first spiral arm (or the total number of stars in the no-spiral-arms model to the 5000 stars chosen for each simulation). Three different values were tried, scaling the values in steps of 25\,\% of the total number of simulated stars. $S_{back}$ refers to the scaling factor of the background population (i.e. stars homogeneously distributed from 100 to 5000\,pc with increasing extinction, see main text) with respect to the total number of stars belonging to the spiral arms. $d_i$ and $A_{H_i}$ indicate the distance and extinction to the $i-th$ spiral arm. The distances and extinctions are given in units of pc and mag, respectively.

 \end{table}

\subsection{Properties of the spiral arms}
\label{comple}

We computed the reduced $\chi^2$ ($\chi^2_{red}$), considering the fixed parameters for each of the MC samples. The minimum $\chi^2_{red}$ values from the MC runs show a Gaussian distribution. We obtained a mean $\chi^2_{red} = 2.43\pm0.01$, where the uncertainty corresponds to the standard deviation of the values.

   \begin{figure*}[h!]
   \includegraphics[width=\linewidth]{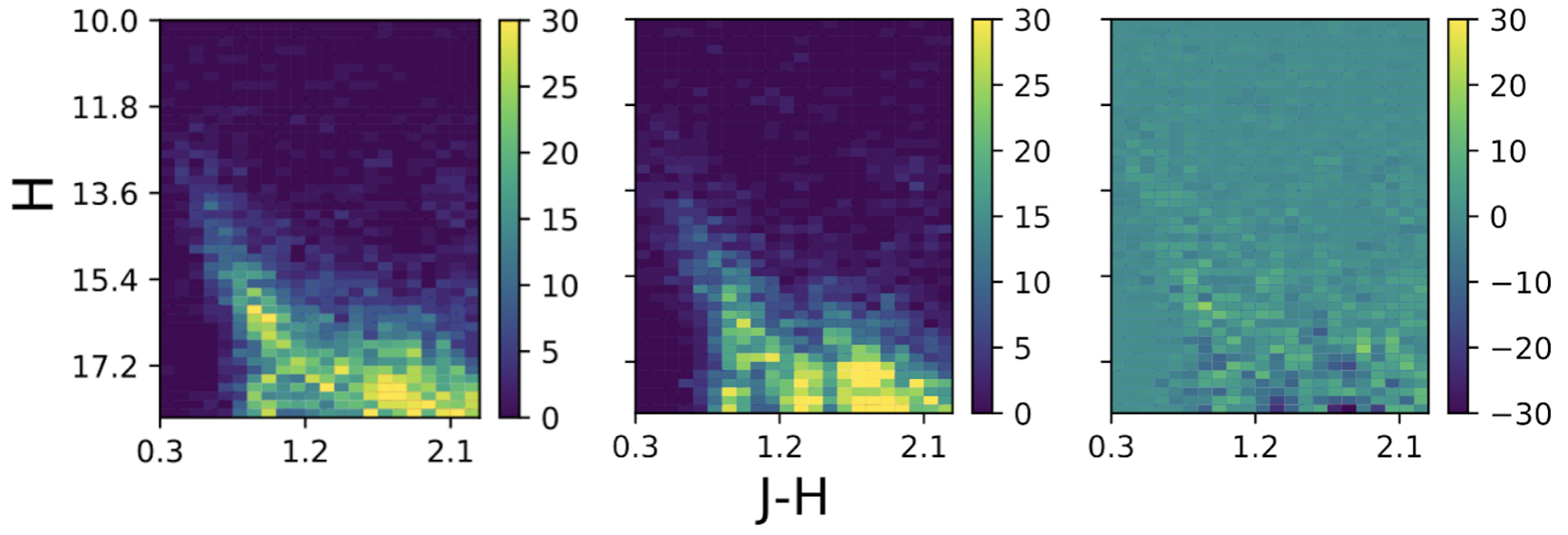}
   \caption{Model fitting performance for a given MC sample (left panel). Middle panel: Best fit for the MC sample ($\chi^2_{red} = 2.24$). Right panel:  Residuals after subtracting the model from the data. The colour bar indicates the number of stars per bin using a linear scale.}

       \label{models}
    \end{figure*}

Table \ref{model_results} shows the best solutions for the parameters fixed averaging over the results obtained for the 500 MC realisations. The uncertainties were obtained considering the maximum between the standard deviation of the parameter distribution for each of the 500 MC samples and the half step of the grid (175 pc and 0.08 mag, for the distance and the extinction, respectively).

\begin{table*}
\caption{Model fitting results.}

\label{model_results} 
\begin{center}
\def\arraystretch{1.4}
\setlength{\tabcolsep}{3.8pt}

\begin{tabular}{cccc}
 &  &  & \tabularnewline
\hline 
\hline 
$JH$  & \multicolumn{1}{c|}{} & $HK_s$  & \tabularnewline
d (pc) & \multicolumn{1}{c|}{$A_H$ (mag)} & d (pc) & $A_{K_s}$ (mag)\tabularnewline
\hline 
1623 $\pm$ 175 & \multicolumn{1}{c|}{0.35 $\pm$ 0.08} & 1685 $\pm$ 259 & 0.11 $\pm$ 0.04\tabularnewline
2641 $\pm$ 178 & \multicolumn{1}{c|}{0.77 $\pm$ 0.08} & 2521 $\pm$ 205 & 0.41 $\pm$ 0.04\tabularnewline
3868 $\pm$ 288 & \multicolumn{1}{c|}{1.68 $\pm$ 0.08} & 3420 $\pm$ 175 & 0.90 $\pm$ 0.04\tabularnewline
4539 $\pm$ 198 & \multicolumn{1}{c|}{2.30 $\pm$ 0.08} & 4931 $\pm$ 272 & 1.19 $\pm$0.04\tabularnewline
\hline 
$S_{back}$ = 0.80 $\pm$ 0.22 & \multicolumn{1}{c|}{} & $S_{back}$ = 0.68 $\pm$ 0.20 & \tabularnewline
\hline 
 &  &  & \tabularnewline
\end{tabular}

\end{center}

\textbf{Notes.} The table includes the results corresponding to the CMDs $H$ vs. $J-H$ and $K_s$ vs. $H-K_s$. The $S_{back}$ parameter refers to the scaling factor of the background population (see Table \ref{model_table}). $d$ and $A_{H}$ indicate the distance and extinction to each of the spiral arms.

 \end{table*}

We addressed the following sources of systematic errors:

\begin{itemize}

\item We analysed the CMD $K_s$ vs. $H-K_s$. We used the synthetic population described in Sect. \ref{mod} and kept the same distance parameters in steps of 350 pc to sample the whole parameter space. We defined a grid of $A_{K_s}$ in agreement with  the $A_H$ grid used for the analysis of the CMD $H$ versus $J-H$. In this way, we created the models assuming four extinction values for each spiral arm varying from 0.075 to 1.2\,mag in steps of 0.075\,mag. We also considered the lower $K_s$-band completeness (mainly due to crowding) in comparison to the $H$ and $J$ bands used previously \citep[80\,\% completeness at $K_s\sim16.3$\,mag, see ][]{Nogueras-Lara:2020aa}. For this, we applied a completeness correction selecting a reference level with a completeness of 50\,\% ($K_s=18.3$\,mag) and randomly removing stars from levels of completeness above in steps of 1\,\%, as described in Sect. 3.1.2 of \citet{Nogueras-Lara:2020aa}. We selected stars fulfilling $K_s\in[11.5, 18.3]$ and $H-K_s\in[0,1.05]$ to account for the significant saturation in $K_s$ for stars brighter than 11.5\,mag \citep{Nogueras-Lara:2019aa} and also to minimise the contamination from the inner Galactic bulge (see Fig.\,\ref{CMD}). We then sampled the data to create the 500 MC realisations in order to apply the model fitting. Table \ref{model_results} shows the results. We concluded that there is no significant variation in the distance parameters within the uncertainties with respect to the results obtained for the $J-H$ case.\\

\item We also analysed the systematic errors due to the chosen bin width. For this, we repeated the fit varying the bin size of the CMD ($\pm 25\%$). The results are presented in Table \ref{app_bins} in Appendix \ref{append1}. We did not observe any significant difference within the uncertainties.\\

\item We tested the influence of the ZP variability on the results. We repeated the fit adding and subtracting the ZP uncertainty (0.04\,mag for $J$ and $H$ bands) independently for each band. The results are presented in Table \ref{app_ZP} in Appendix \ref{append1} and agree well with our results.\\

\item To study the influence of the selected models, we repeated the analysis using BaSTI models \citep{Pietrinferni:2004aa,Pietrinferni:2006aa} to generate the synthetic population described in Sect. \ref{mod}. Because the synthetic population generated using the BaSTI  web tool\footnote{http://basti.oa-teramo.inaf.it} only considers stars with masses larger than 0.5\,$M_\odot$, we restricted the lower limit of the $H$ cut to 17\,mag instead of 18.3 mag to avoid gaps when analysing the closest spiral arms. The results are presented in Table \ref{app_basti} in Appendix \ref{append1}. We did not observe any significant difference within the uncertainties.\\

\item We also tested the possible systematic errors introduced by the chosen SFH. For this, we used a BaSTI \citep{Pietrinferni:2004aa,Pietrinferni:2006aa} model that considers a stellar population similar to that expected for the local disc according to \citet{Rocha-Pinto:2000aa} as it is implemented in the BaSTI web tool. The selection of the stars was also limited to $H=17$\,mag given the limitations of the BaSTI models. The results are presented in Table \ref{app_basti} in Appendix \ref{append1}. We did not observe any significant difference within the uncertainties.\\

\item Finally, to check the method, we generated three simulated data sets considering the same SFH but more realistic situations. We varied the relative number of stars between arms, the width of the spiral arms, and the relative contribution of the stellar background. We applied the method described previously,  now using 100 MC samples. The results are presented in Table \ref{app_tests} of Appendix \ref{append1}. We find that the extinction of all the features is always determined within the uncertainties in all the cases. On the other hand, we measured some deviation in the results for the distances of the first simulated spiral arm. The determination of the distance is more dependent on the precise shape of the spiral arm features given that it only affects the $y$-axis (the colour, $x$-axis, does not depend on distance). Thus, it is more influenced by the scatter of the data, or by a less defined spiral arm feature. In this sense, the determination of the distance for the first spiral arm might be more affected because its number of stars is lower in comparison with the other spiral arms. Nevertheless, the measured deviations are always within 1.5$\,\sigma$. In the case of real data, we believe that the distance to the first spiral arm is properly determined given the agreement with Gaia (Sect. \ref{gaia}) data and also with previous work. It might be possible that the scatter of the data is overestimated when creating the simulations. We also measured some deviation when computing the influence of the background population. Nevertheless this is expected given our simple approach, which is designed to  simply identify the extinction and distance to the spiral arms.\\

\end{itemize}

\subsection{The role of extinction}
\label{ext_fit_model}

The extreme extinction towards the GC, in particular towards the NB \citep[e.g.][]{Nogueras-Lara:2019ad}, allows us to clearly distinguish the foreground population and trace the spiral arms as shown above. Nevertheless, some contamination from the bulge is expected for the fourth spiral arm, in particular when using the $K_s$ band. We overcame this difficulty when analysing the $J$ and $H$ bands, where the effect of extinction is larger \citep[e.g.][]{Nishiyama:2009oj}, making it easier to distinguish between components with different reddening. Comparing the results with the ones obtained when applying the technique to the CMD $K_s$ versus $H-K_s$, allows us to assess the consistency between the distance and extinction results. We ended up with a larger distance for the fourth spiral arm when using the CMD $K_s$ versus $H-K_s$, although this result remains compatible within the uncertainties with the results from the CMD $H$ versus $J-H$ (Table\,\ref{model_results}). Therefore, we conclude that the obtained parameters are not significantly affected by this contamination.

A similar analysis is not possible at higher latitudes given the lower extinction towards the Galactic bulge at these latitudes. The inner bulge fields analysed in Sect. \ref{gaia} correspond to an average extinction of $A_{K_s}\sim1.3$\,mag \citep[averaging over the values obtained by][for two inner-bulge regions located at $\sim 0.6^\circ$ and $\sim 0.4^\circ$ to the Galactic north of the Milky Way centre]{Nogueras-Lara:2018ab}, which is only about half of the total extinction towards the central field (latitude $\sim 0^\circ$).  Moreover, the larger size of the central field and the lower density of stars given the projection effects for higher latitudes also affect the observed number of stars from the spiral arms in the transition and the inner bulge regions. 

\begin{figure}
   \includegraphics[width=\linewidth]{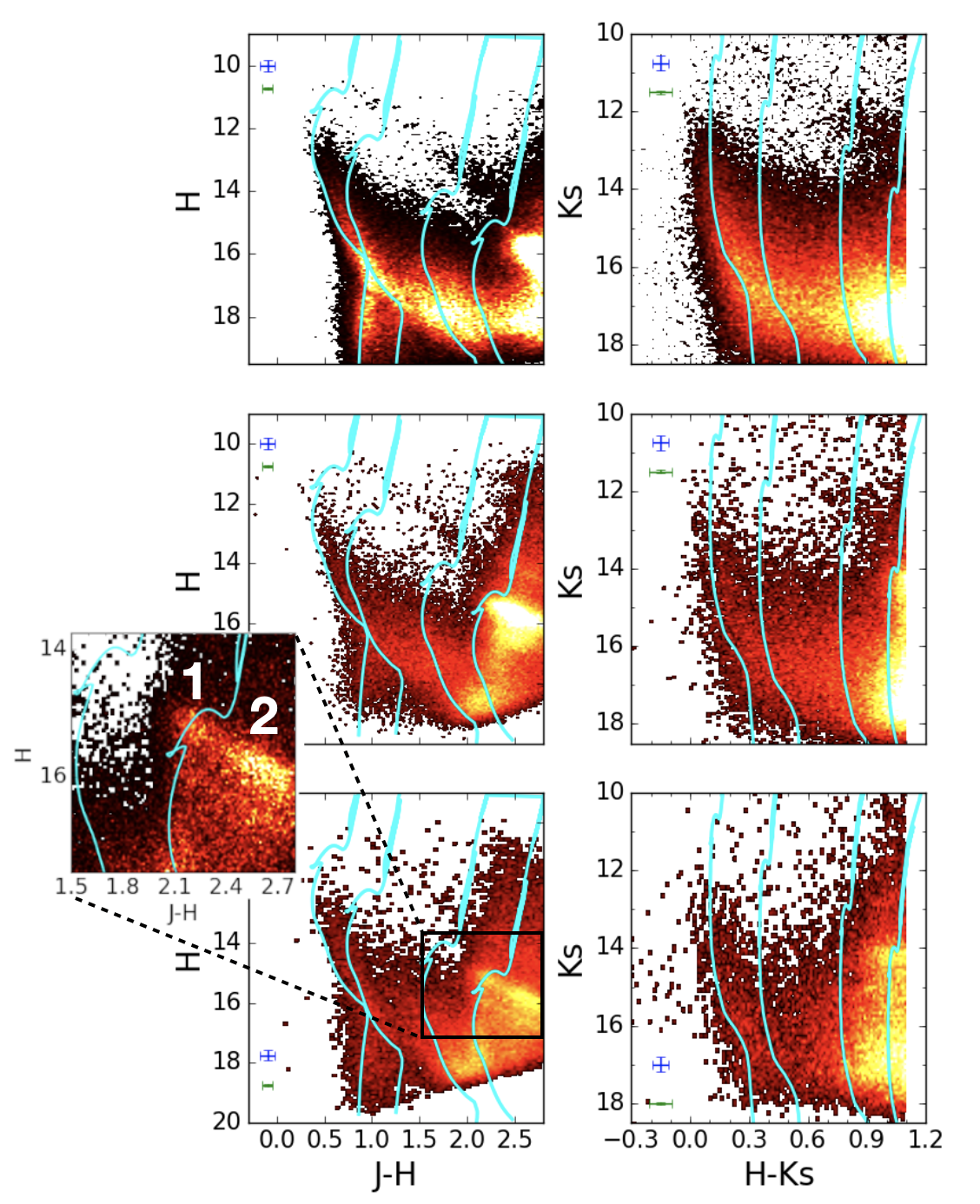}
   \caption{CMDs $H$ vs. $J-H$ and $K_s$ vs. $H-K_s$ for the central, transition, and inner bulge regions of the GALACTICNUCLEUS survey, from the upper to the lower panels, respectively. The colour code refers to stellar densities, using a power-stretch scale (different in each case to stress the spiral arm features). Cyan lines correspond to a PARSEC isochrone of 1\,Gyr and $[M/H = -0.07]$ according to the extinction and distance parameters obtained for the spiral arm. The blue and green error bars indicate the 1\,$\sigma$ uncertainties of the position of the isochrones and the photometric ZP, respectively. The zoomed-in region corresponds to the red clump features \citep[e.g.][]{Girardi:2016fk} from the inner Galactic bulge (`1' and `2' associated to different extinctions from the IBS and IBS regions) and the overlap with the fourth spiral arm.}

       \label{density_2x3}
    \end{figure}

Therefore, the stars in the third and fourth (and to a certain degree even the second) spiral arm from the Earth overlap with the GC stars in the CMDs for the bulge fields. Figure\,\ref{density_2x3} illustrates this effect showing the CMDs $H$ versus $J-H$ and $K_s$ versus $H-K_s$ for the studied regions. We over-plotted PARSEC isochrones of 1\,Gyr and $[M/H] = -0.07$ according to the parameters of distance and extinction obtained for the $J-H$ and $H-K_s$ data of the central region (Table \ref{model_results}). We observed good agreement between the position of the isochrones and the over-densities in the CMDs, within the uncertainties. Given the lower extinction, the features corresponding to the GC are less extinguished for the transition and the inner bulge regions than for the central one. In this sense, the middle and lower panels of Fig.\,\ref{density_2x3} show how the RC features from the inner bulge or the GC stellar population are contaminated by the presence of stars from the main sequence belonging to the fourth spiral arm. This effect is visible in the CMDs $H$ versus $J-H$, where the RC feature shows an excess of stars belonging to the isochrone associated to that spiral arm. The zoomed-in region in Fig. 4, corresponding to the inner bulge regions, depicts this effect. In this particular case, the different extinction from the IBN and IBS \citep{Nogueras-Lara:2018ab} magnifies the overlap.

\subsection{Discussion}
\label{distances_grid}

A detailed two-dimensional fit of the experimental CMDs is very complex given the high number of free variables to be considered. Namely, the presence of stars with different ages and/or metallicities, the different contributions from each spiral arm depending on its corresponding distance and/or width, the differential extinction, and the possible contamination from the GC stellar population for the innermost arms contribute to this increase in complexity. 

Nevertheless, we did not aim to characterise the stellar population and/or metallicities of the analysed stars, but to compute the distance to the spiral arms and their extinction given the average features in the CMDs. We demonstrate that our method is good enough for our purposes, addressing potential sources of systematic errors and trying different stellar populations and models. Our results agree well with the distances from the recent work by \citet{Reid:2019aa}. From  Fig. 1 of this latter paper we obtained $d_1\sim1.4$, $d_2\sim2.7$, $d_3\sim3.7$, and $d_4\sim4.7$ kpc, which are fully compatible with our results.

The obtained structures would correspond to the Sagittarius–Carina, the Scutum–Centaurus–OSC, the Norma–Outer, and the 3 kpc arms. As reference values for the distance towards the spiral arms, we used the results obtained for the CMD $H$ versus $J-H$ because of the lower contamination from the GC and also because the shape of the isochrones is more sensitive to changes in the $y$-axis as can be seen in Fig.\,\ref{density_2x3}.

\section{Analysis of Gaia sources}
\label{gaia}

To cross-check the obtained distances to the spiral arms, we cross-correlated the regions from the GALACTICNUCLEUS survey with the Gaia DR2 catalogue \citep{Gaia-Collaboration:2018aa}. The mean offset between the two catalogues is $\Delta$RA = $0.05\pm0.02$ arcsec and $\Delta$Dec = $0.04\pm0.02$ arcsec. We defined a maximum distance of $\sim0.15''$  to
cross-correlate the catalogues. Moreover, we only accepted stars whose distances was measured with more than 3\,$\sigma$ significance, and removed all the stars with distances larger than 10 000\,pc. In this way, we avoid erroneous identifications and stars with spurious astrometric solutions \citep{Gaia-Collaboration:2018aa}.

\begin{figure}
   \includegraphics[width=\linewidth]{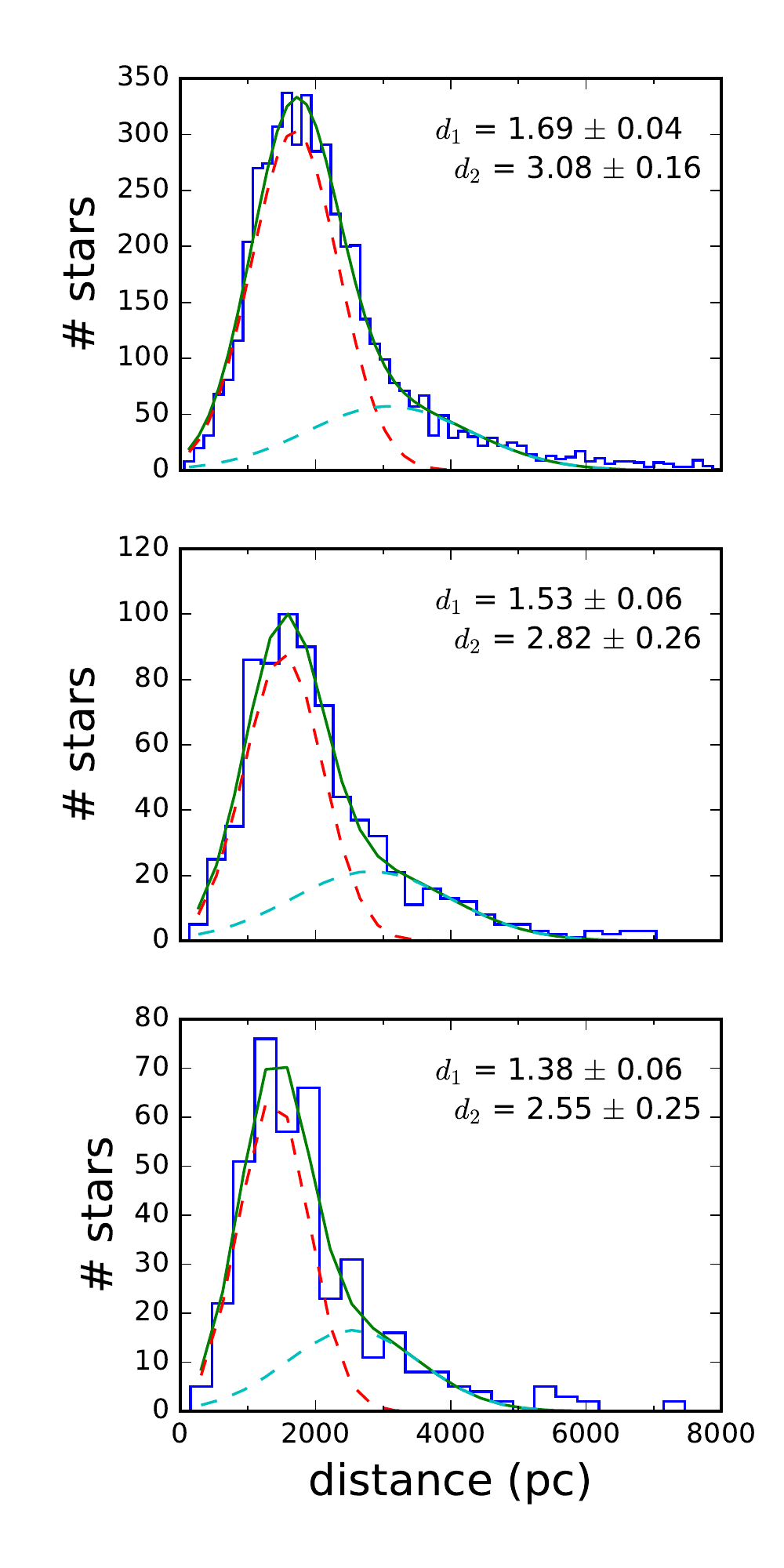}
   \caption{Histograms of the distances to the common stars between Gaia and GALACTICNUCLEUS detected in the central, transition, and inner bulge regions (upper, middle, and lower panel, respectively). The solid green lines indicate the two-Gaussian fits, whereas the red and cyan dashed lines show each of the two Gaussian components. The mean values and their errors for each component are indicated for each panel in units of kpc.}

       \label{GAIA_distance}
    \end{figure}

The high extinction towards the GC and the optical photometry from Gaia only allowed us to obtain common stars whose colours are restricted to $J-H\lesssim1$ mag for faint stars ($H\lesssim16$ mag) and $J-H\lesssim1.5$ mag for some bright stars around $H\lesssim12$\,mag. To obtain the distances to the detected common stars, we used the Gaia parallaxes taking into account that, in general, they cannot be directly inverted to get the distance to a given source \citep[e.g.][]{Luri:2018aa,Bailer-Jones:2018aa}. We applied the near-linear trend of distance bias ($\delta_p = -0.054$) derived by \citet{Schonrich:2019aa}. To remove any additional problem related to the distance to  individual stars, we computed average distances building histograms of the underlying distributions corresponding to each of the analysed regions (Fig. \ref{GAIA_distance}). We ended up with $\sim 5000$, $\sim 800$, and $\sim 400$ stars for the central, transition, and inner bulge regions, respectively.

Given the high extinction along the line of sight towards the GC, we can only detect stars located at $\lesssim 3$kpc from Earth using Gaia data for this line of sight, restricting the analysis to the closest spiral arms. We detected a bimodal distribution that can be fitted well with a two-Gaussian model. We computed the average distance to each of the features as the mean value of each of the Gaussians. The uncertainties were estimated considering the error of the mean. The results and the associated uncertainties are over-plotted in Fig. \ref{GAIA_distance}. We find that the two-Gaussian distribution fits the data better than a single Gaussian when selecting stars with distances of < 5000\,pc and using the SCIKIT-LEARN Python function {\it GaussianMixture} \citep[GMM][]{Pedregosa:2011aa} to compute the Bayesian information criterion \citep{Schwarz:1978aa} and the Akaike information criterion \citep{Akaike:1974aa}. We find that the two-Gaussian distribution is preferred. We did not try any more complex models because of the low completeness of the data beyond 3000\,pc, and so no more components are expected (nor visually identified) in the distributions.

Our results agree well with the presence of two spiral arms, particularly in the case of the central and the transition regions, where the number of stars is larger than for the inner bulge, and where the distance values obtained are somewhat smaller. The secondary Gaussian feature is significantly smaller than the first one, as expected. This is because only a small fraction of the stars belonging to the second spiral arm were detected in the Gaia survey given the extinction and the larger distance.  We computed a mean value for the distance to each of the features averaging over the obtained results and ended up with $d_{1\_Gaia} = 1.5\pm0.1$\,kpc and $d_{2\_Gaia}  = 2.8\pm0.2$\,kpc. The uncertainties refer to the standard deviation of the measurements towards the three different regions. The obtained results agree within the uncertainties with the distances obtained for the first and the second spiral arms in the previous section and in the literature (see Sect. \ref{intro_sect}).

\section{Analysis of the colour--colour diagram}

Figure \ref{ccd} shows the colour--colour diagram (CCD) $J-H$ versus $H-K_s$ of the central region of the GALACTICNUCLEUS survey. The highest density region in the CCD ($H-K_s>1.3$ mag and $J-H>2.65$\,mag) corresponds to the GC. These are stars from the NB and the inner Galactic bulge.  Excluding these, there are two main over-density features ($J-H\sim1$\,mag and $J-H\sim1.8$\,mag) that can be identified in the right panel of Fig. \ref{ccd}. According to the analyses carried out in the previous sections, these overdensities correspond to the spiral arm structure along the line of sight from the Earth to the GC. We built histograms of the $J-H$ and $H-K_s$ distributions to further analyse these features (right panels in Fig. \ref{ccd}):

   \begin{figure*}
   \includegraphics[width=\linewidth]{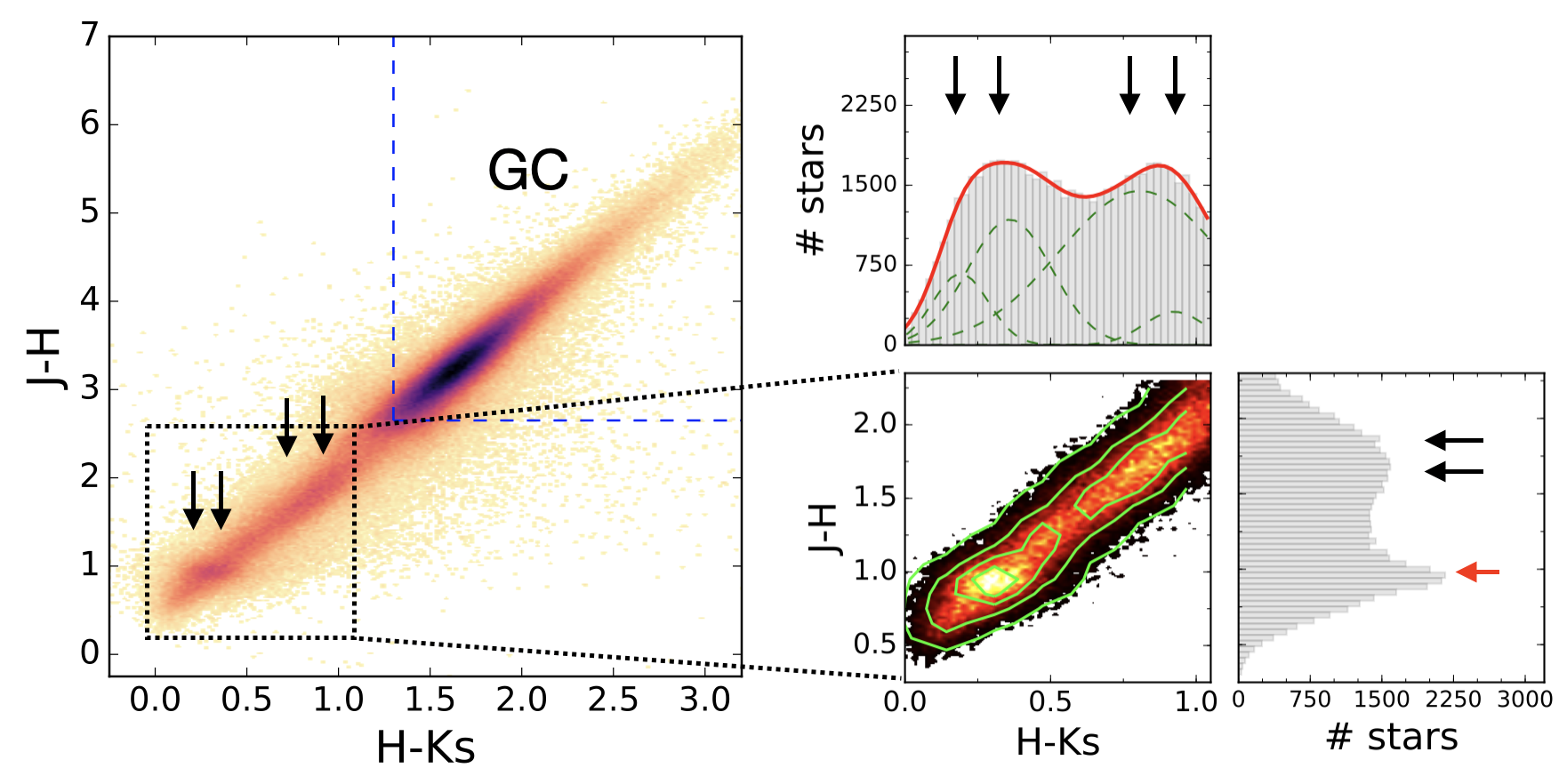}
   \caption{Left panel: Colour--colour diagram $J-H$ vs. $H-K_s$ corresponding to the central region of the GALACTICNUCLEUS survey. The black arrows indicate the over-density features that correspond to the spiral arms detected in Sect. \ref{cmd_fitting}. The blue dashed lines mark the stellar population belonging to the GC, where the density is significantly different from the rest. Right panel: Zoom into the foreground population from the spiral arms and histograms corresponding to the underlying distributions. The red solid line indicates the result of a four-Gaussian fit (green dashed lines for each individual Gaussian). The black arrows show the mean location of the spiral arms. The red arrow indicates the overlap between the first and the second spiral arms in one single feature in the histogram $J-H$.}

       \label{ccd}
    \end{figure*}

\begin{itemize}

\item $J-H$: Given the overlap between the isochrones corresponding to the first and the second spiral arms that we observed in the CMDs $H$ versus $J-H$ (left panels  Fig. \ref{density_2x3}), it is not possible to distinguish the four spiral arms in the histogram corresponding to the $J-H$ distribution. We observed a prominent narrow peak around $J-H\sim1$\,mag and a secondary extended peak around $J-H\sim1.75$\,mag. According to the synthetic CMD fitting, the first peak is probably the result of the overlap between the first and the second spiral arms. On the other hand, the extended peak might be the result of the overlap between the third and the fourth spiral arms. \\

\item $H-K_s$: There are two clear features that probably correspond to the overlap between the first and second, and third and fourth spiral arms. Given that the overlap between the  isochrones of the spiral arms is not as significant as for $J-H$ (see Fig.\,\ref{density_2x3}), we tried to fit this histogram using a four-Gaussian model to check whether it is compatible with the structure obtained previously. This is a simplistic approach that only attempts to detect over-densities associated to the spiral arms in the CCD. 

We used the stars detected in all three bands ($JHK_s$) within the same magnitude ranges used in Sect. \ref{cmd_fitting} for the analysis of the CMDs $H$ versus $J-H$ and $K_s$ versus $H-K_s$. We fitted the model using the Python SCIPY routine {\it curve-fit} and the initial values according to the expected positions of the spiral arm features. We only imposed that the four Gaussians cannot take negative values for the fit. We obtained that this simple model agrees well with the results from previous sections. We obtained mean values of $(H-K_s)_1 \sim 0.2$, $(H-K_s)_2 \sim 0.4$, $(H-K_s)_3 \sim 0.8$, and $(H-K_s)_4 \sim 0.9$ mag, where the subindex indicates the number of each spiral arm. The feature corresponding to the fourth spiral arm is smaller than the others. Nevertheless, this is expected given the colour cuts applied to limit possible confusion with the stars belonging to the inner Galactic bulge, whose extinction is close to that from this spiral arm \citep{Nogueras-Lara:2018ab}.

\end{itemize}

Therefore, the analysis of the CCD allows us to distinguish between GC and spiral arm stellar populations and agrees with the structure derived previously.

\section{Extinction}

According to the previous analysis, we find that the CMD $K_s$ versus $H-K_s$ is the best choice to select the reference stars with which to analyse the extinction corresponding to each feature. This is because the stellar population belonging to each spiral arm can be disentangled more easily there than in the CMD using the $J$ band, where the isochrones from different spiral arms overlap (see Fig.\,\ref{density_2x3}).

We only used the central region to analyse the extinction for two reasons: (1) The average extinction of the GC stellar population is maximum for this region, allowing us to detect the innermost spiral arms. (2) The number of available reference stars is largest given the covered area, which implies that more stars are available to analyse the extinction.

\begin{figure}
   \includegraphics[width=\linewidth]{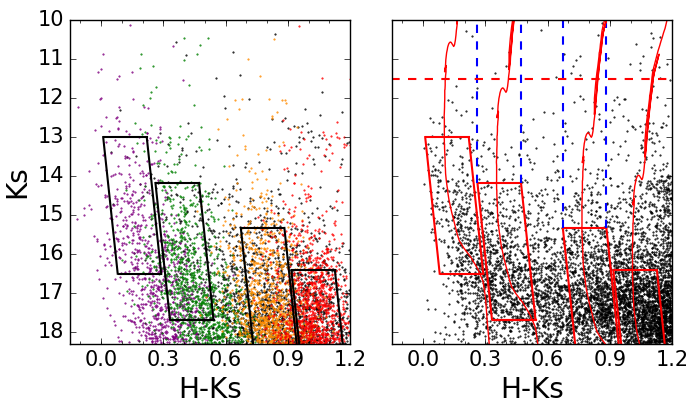}
   \caption{Colour--magnitude diagrams $K_s$ vs. $H-K_s$. Left panel: Best model-fit solution obtained for the CMD $K_s$ vs. $H-K_s$ (see Table \ref{model_results}). Different colours indicate the stars contributing from each of the spiral arms. In particular, purple, green, orange, red, and black dots indicate stars from the first, second, third, and fourth spiral arms, and background sources respectively. Black solid parallelograms mark the selection of reference (mainly main-sequence) stars to analyse the extinction corresponding to each of the spiral arms. Right panel: Real data considering completeness correction as indicated in Sect. \ref{comple}. The red dashed line indicates the saturation limit of the $K_s$ photometry ($K_s = 11.5$\,mag). The red solid parallelograms depict the reference main sequence stars according to the best model fit as shown in the left panel. The isochrones corresponding to a stellar population of 1\,Gyr and $[M/H] = -0.07$ are over-plotted in red (see Sect. \ref{comple}). The blue dashed lines indicate the stars added to the reference main sequence stars to create a $K_s$ luminosity function for the second and third features (see Sect. \ref{KLF}).}

       \label{ext_maps_selection}
    \end{figure}

To select the reference stars belonging to each of the spiral arms, we generated a synthetic stellar population according to the best fit obtained for the CMD $K_s$ versus $H-K_s$ (see Sect. \ref{comple}). The synthetic model indicates the regions where the probability of finding stars from a given spiral arm is maximum. In this way, we defined a selection box to identify them in the real data (left panel of Fig. \ref{ext_maps_selection} ). Figure \ref{ext_maps_selection} shows the chosen stars with which  we analyse the extinction.

\subsection{Extinction curve}
\label{extinction_parsec}

We computed the ratios between extinctions $A_J/A_H$ and $A_H/A_{K_s}$ using the selected stars belonging to the spiral arms, and then used these ratios to derive the extinction curve. To select the reference stars in the $J$ band avoiding the overlap of the spiral arm features (see Fig. \ref{ext_maps_selection}), we used the reference stars in the CMD $K_s$ versus $H-K_s$ that were also detected in $J$. To calculate the extinctions $A_i$ (where $i$ indicates the photometric band), we computed the intrinsic colours ($J-H$ and $H-K_s$) of each of the reference stars interpolating from a PARSEC isochrone (Fig. \ref{ext_maps_selection}) corresponding to the best-fit parameters obtained in Sect. \ref{cmd_fitting}.

Given the completeness due to crowding of the $K_s$-band data, we applied a completeness correction as explained in Sect. \ref{comple}. As this approach is based on the removal of stars above a reference completeness limit of 50\,\%, we generated 100 MC samples, randomly removing stars to correct for completeness (see Sect. 3.1.2 of \citet{Nogueras-Lara:2020aa} for further details). We computed the ratios between extinctions following the equation:

\begin{equation}
\label{ratio}
A_{i}/A_{j} =\frac{i-j-(i-j)_0}{ext_{j}}+1  \hspace{0.5cm},
\end{equation}

\noindent where $i$ and $j$ are the photometric bands, the subindex $`0$' indicates intrinsic colour and $ext_j$ is the mean extinction to a spiral arm as indicated in Table \ref{model_results}. We averaged over all the reference stars for a given spiral arm and computed the uncertainties quadratically propagating the uncertainties corresponding to the magnitudes involved in Eq. \ref{ratio}. In particular, we considered the ZP systematic uncertainty of 0.04 mag for each band in the calculation, and the uncertainty of the mean extinction to a given spiral arm. 

We obtained the final values and uncertainties as the mean of the individual results of each of the 100 MC iterations. Table \ref{ratios} summarises the results. We find that the extinction ratios $A_J/A_H$ and $A_H/A_{K_s}$ do not vary for the different spiral arms considered, within the uncertainties. We used a weighted mean to compute mean values of the ratios taking into account the larger uncertainties obtained for the first spiral arm, where the lower number of stars and the possible contamination by  stars close to the Sun might influence the results. We ended up with $A_J/A_{H} = 1.89 \pm 0.11$ and $A_H/A_{K_s} = 1.86 \pm 0.11$.  These values are in agreement with the recent results obtained for the GC ($A_J/A_H = 1.87\pm0.03$ and $A_{H}/A_{K_s} = 1.84\pm0.03$) by \citet{Nogueras-Lara:2019ac,Nogueras-Lara:2020aa}. Therefore, our results point towards a more complex extinction curve in the NIR than the simple power-law approach that is generally accepted \citep[e.g.][]{Nishiyama:2006tx}. This more complex behaviour of the extinction curve might explain the different values obtained in the literature \citep[e.g][]{Nishiyama:2009oj,Stead:2009uq,Fritz:2011fk,Alonso-Garcia:2017aa}, as discussed in Sect.\,7 in \citet{Nogueras-Lara:2020aa}.

These values are in agreement with the results obtained for the GC ($A_J/A_H = 1.87\pm0.03$ and $A_{H}/A_{K_s} = 1.84\pm0.03$) by \citet{Nogueras-Lara:2020aa}. The constant extinction ratios for stars belonging to different structures and distances indicate that the extinction curve in the NIR bands $JHK_s$ towards the GC does not vary significantly with distance. Therefore, the extinction law obtained for the GC \citep[e.g. ][]{Nogueras-Lara:2019ac,Nogueras-Lara:2020aa} is also applicable when studying the foreground population corresponding to the spiral arm structure from the Earth to the GC.

\begin{table}
\caption{Extinction curve using spiral arm reference stars.}
\label{ratios} 
\begin{center}
\def\arraystretch{1.4}
\setlength{\tabcolsep}{3.8pt}

\begin{tabular}{ccc}
 &  & \tabularnewline
\hline 
\hline 
\# spiral arm & $A_J/A_H$ & $A_H/A_{K_s}$\tabularnewline
\hline 
1 & 2.07$\pm$0.31 & 1.94$\pm$0.39\tabularnewline
2 & 1.89$\pm$0.21 & 1.83$\pm$0.21\tabularnewline
3 & 1.89$\pm$0.19 & 1.86$\pm$0.19\tabularnewline
4 & 1.82$\pm$0.19 & 1.87$\pm$0.19\tabularnewline
\hline 
Mean value & 1.89$\pm$0.11 & 1.86$\pm$0.11\tabularnewline
\hline 
 &  & \tabularnewline
\end{tabular}
\end{center}

\textbf{Notes.} The mean values in the last row correspond to a weighted mean of the extinction ratios obtained for the spiral arms in the rows above.

\end{table}

\subsection{Extinction maps}
\label{extinction_m}

Once we had determined the spiral arm structure, we built $K_s$-extinction maps corresponding to each of the spiral arms in order to further analyse how the extinction varies along the line of sight towards the GC. We used all the available stars within the selection box without applying the completeness correction. This is because detecting less stars due to completeness problems will not influence the values of the extinction. On the contrary, using all the stars allows us to increase the number of reference stars to improve the quality of the extinction maps.

\subsubsection{Methodology}

We built the extinction maps using stars within the reference boxes shown in Fig. \ref{ext_maps_selection}. We converted the observed stellar colour $H-K_s$ into extinction following the equation:

\begin{equation}
\label{eq_maps}
ext_{K_s} = \frac{H-K_s-(H-K_s)_0}{A_H/A_{K_s}-1} \hspace{0.5cm},
\end{equation}

\vspace{0.2cm}

\noindent where $H$ and $K_s$ are the observed magnitudes, $A_H/A_{K_s}$ is the relation between extinctions obtained in \citet{Nogueras-Lara:2020aa} and is equal to $1.84\pm0.03$, and $(H-K_s)_0$ refers to the intrinsic colour, computed for each star interpolating from a PARSEC isochrone as explained in Sect. \ref{extinction_parsec}. Equation \ref{eq_maps} was adapted from Eq. 5 in \citet{Nogueras-Lara:2018aa} to avoid using the effective wavelengths, given that they vary for different stellar types and extinctions \citep[e.g.][]{Nogueras-Lara:2020aa}.  

To create the extinction maps, we applied the approach described in \citet{Nogueras-Lara:2018aa,Nogueras-Lara:2019ad}, but using the reference stars previously specified instead of RC stars. Namely, we defined a pixel size of $20''$ (larger than in the \citet{Nogueras-Lara:2018aa,Nogueras-Lara:2019ad} given the significantly lower number of reference stars, and the significantly lower influence of the differential extinction for the spiral arm than for the GC stellar population), and computed the extinction for each pixel using at least five reference stars within a radius of $30''$ from the centre of each pixel. We also applied an inverse distance weight method ($p=0.25$) to consider the different distances of each reference star to the centre of the pixel (see Sect. 7 of \citet{Nogueras-Lara:2018aa} for further details). We computed the statistical uncertainties considering the variation of the extinction for a given pixel using a jackknife algorithm. Moreover, we estimated the systematic uncertainty taking into account the uncertainties of $A_H/A_{K_s}$ and the ZP systematic uncertainty \citep[0.04 mag for both $H$ and $K_s$ bands][]{Nogueras-Lara:2019aa}.

\begin{figure*}
   \includegraphics[width=\linewidth]{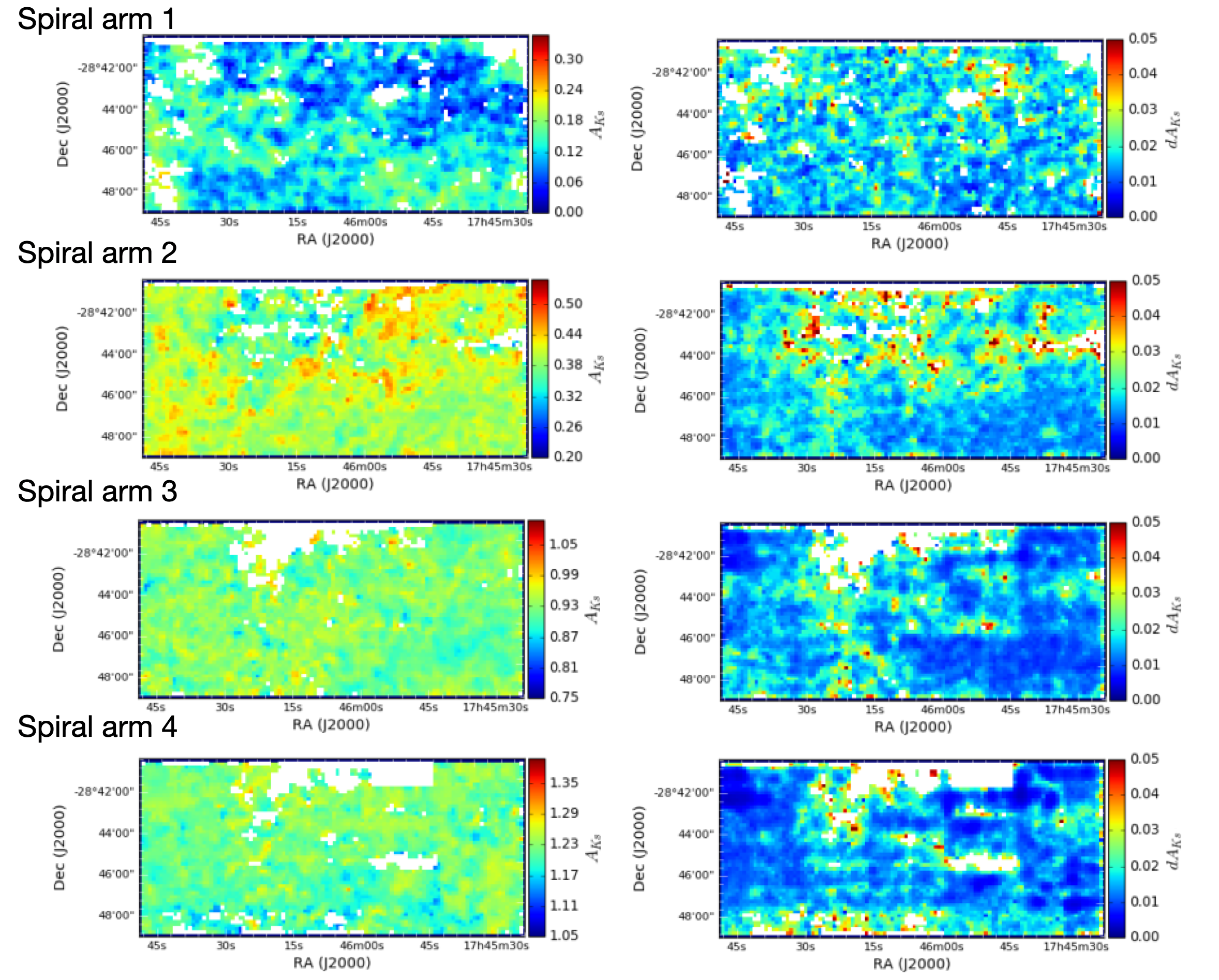}
   \caption{Left column: Extinction maps corresponding to each spiral arm. Right column: Associated uncertainties obtained using a jackknife algorithm (see main text for further details). White pixels indicate regions where the number of reference stars is not enough to compute an extinction value. The white rectangle in the upper part of the fourth spiral arm corresponds to a region of poor quality data in $H$ band \citep[see Table A.1. in ][]{Nogueras-Lara:2019aa}. Given the increase in extinction for further spiral arms, the colour scale is different for each of the extinction maps. The extinction range is always the same, 0.35\,mag, in order to make small extinction variations within the same extinction layer  visible.}

       \label{ext_maps}
    \end{figure*}

Figure \ref{ext_maps} and Table \ref{parameters_maps} show the results. We obtained that the extinction maps for the different spiral arms are quite homogeneous and do not present any significant extinction variations. The largest variations are measured for the first spiral arm, where the lower number of stars and the possible influence of stars in the vicinity of the Sun (introducing large differential distances between the stars within this map) might contaminate our results. Moreover, the lower value of the mean extinction for the first layer makes any relative difference more visible due to the denominator of Eq. \ref{eq_maps}. 

\begin{figure}
   \includegraphics[width=\linewidth]{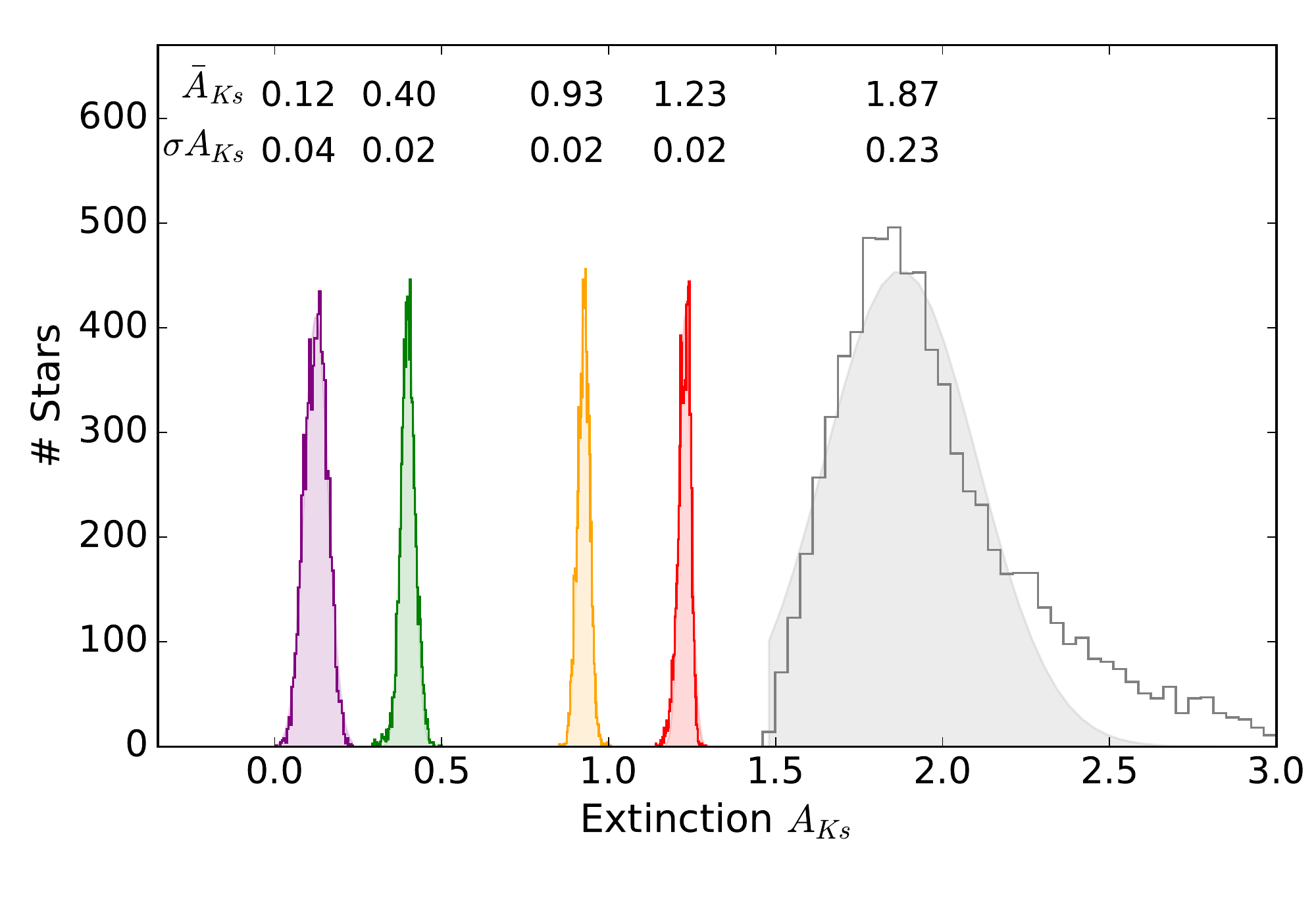}
   \caption{Histograms of the $A_{K_s}$ values per pixel in the extinction maps. The blue, green, orange, and red histograms correspond to the spiral arms. The grey histogram was obtained using the extinction map in \citet{Nogueras-Lara:2019ad}, corresponding to the central region of the NSD. Given the smaller area of the NSD extinction map, the spiral arm histograms were properly scaled. Filled coloured contours show Gaussian fits, whose mean and standard deviation are indicated in the figure.}

       \label{histogram_extinction}
    \end{figure}

To check the homogeneity of the derived extinction layers, Fig.\,\ref{histogram_extinction} shows histograms of the extinction computed for the spiral arm reference stars using the corresponding extinction maps. We obtained that a Gaussian model properly fits the data for all of the spiral arms. The standard deviations of the distributions (over-plotted in Fig. \ref{histogram_extinction}) are similar in all the cases, but are somewhat larger for the first spiral arm as expected given the reasons explained above. We concluded that the extinction is homogeneous within each spiral arm feature and that there is no overlap between them. Therefore, they appear to be independent extinction layers. We also compared them with an extinction map obtained for a region of the NB \citep{Nogueras-Lara:2019ad}. In this case, the Gaussian fit is not adequate given the inhomogeneity and the larger extinction variations within the line of sight. On the other hand, we compared  the mean values obtained in Table\,\ref{parameters_maps} with the results from the analysis of the  CMD $K_s$ versus $H-K_s$ (Table\,\ref{model_results}) and checked that they agree within the uncertainties.

\begin{table}
\caption{Extinction values from the extinction maps.}
\label{parameters_maps} 
\begin{center}
\def\arraystretch{1.4}
\setlength{\tabcolsep}{3.8pt}
    
\begin{tabular}{ccccc}
 &  &  &  & \tabularnewline
\hline 
\hline 
Spiral  & \# reference & $A_{K_s}$ & $\Delta_{stat}$ & $\Delta_{syst}$\tabularnewline
arm & stars & (mag) & (mag) & (mag)\tabularnewline
\hline 
1 & 7756 & 0.12 & 0.02 & 0.06\tabularnewline
2 & 13468 & 0.40 & 0.02 & 0.07\tabularnewline
3 & 19594 & 0.93 & 0.02 & 0.07\tabularnewline
4 & 24405 & 1.23 & 0.01 & 0.09\tabularnewline
\hline 
 &  &  &  & \tabularnewline
\end{tabular}

\end{center}
\textbf{Notes.} Column  \#reference stars indicates the number of reference stars used to create the maps. $A_{K_s}$ indicates the mean value obtained for the stars in the selection boxes in Fig. \ref{ext_maps_selection}. $\Delta_{stat}$ and $\Delta_{syst}$ refer to the statistical and systematic uncertainty.

 \end{table}

\subsubsection{Inter-arm extinction}

\begin{figure}
       \begin{center}
   \includegraphics[width=0.8\linewidth]{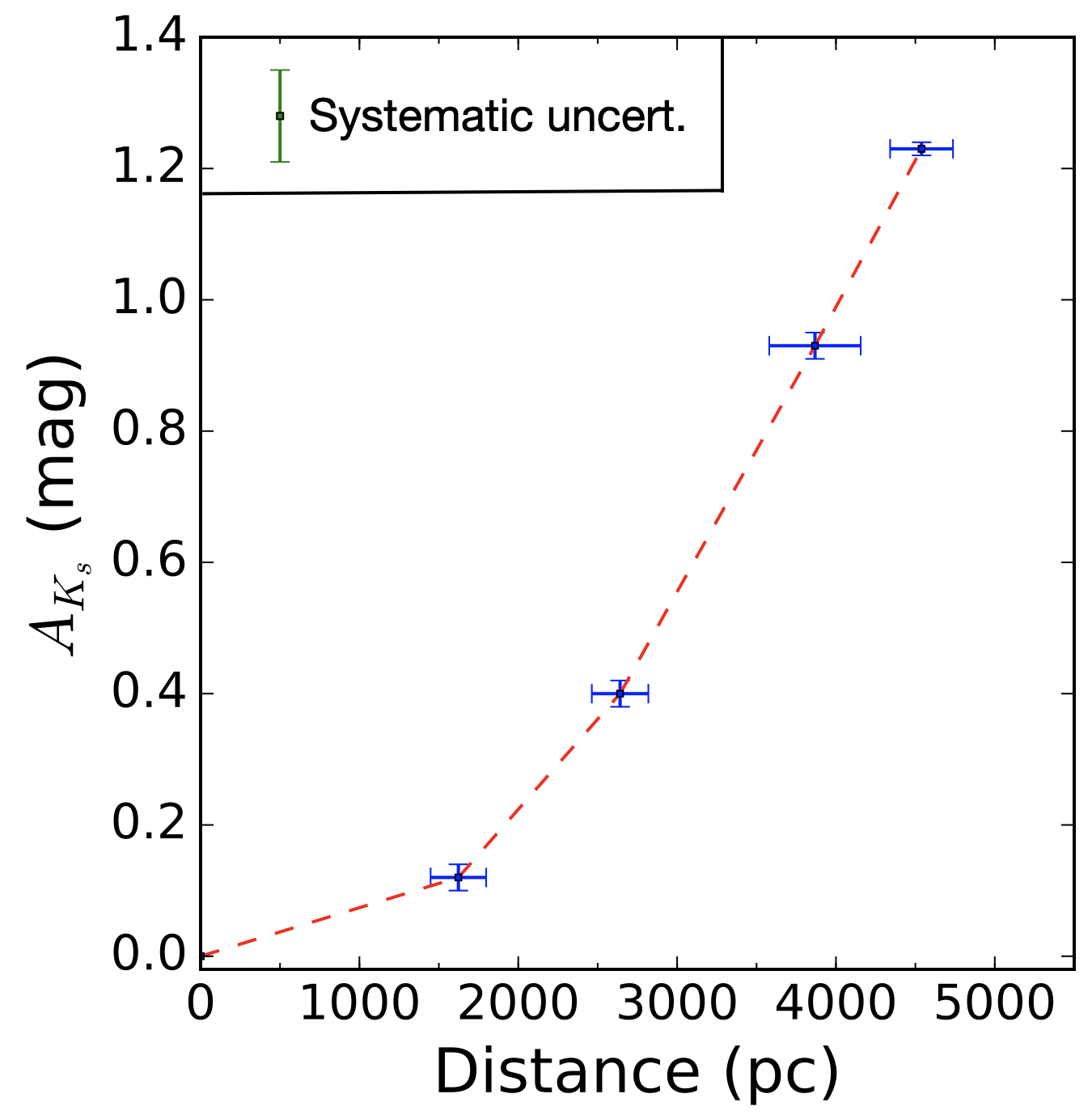}
   \caption{Extinction vs. distance. The blue points correspond to the results obtained for the distance (see Table \ref{model_results}) and extinction $A_{K_s}$ (Table \ref{parameters_maps}) obtained for the spiral arms features. The dashed red line shows a linear fit of the data points, two by two. The green error bar indicates the systematic uncertainty of the extinction, which does not affect the relative variation between arms because all the arms are affected in the same way.}  
  
       \label{relative_ext}
          \end{center}  
    \end{figure}

We computed the differential extinction per kiloparsec ($A_{K_s}/d$) for each of the spiral arms. We used  the results from Table \ref{parameters_maps} and the distances to each of the spiral arms obtained in Sect. \ref{distances_grid}. We find  $0.07\pm0.01$, $0.28\pm0.07$, $0.43\pm0.12$, and $0.45\pm0.24$ mag/kpc for each of the spiral arms, from the closest to the furthest, respectively. The uncertainties were computed quadratically propagating the uncertainties of the $A_{Ks}$ and the distances. As the main goal is to compare the differential extinction per kiloparsec for the spiral arms, we did not consider the systematic uncertainty of $A_{K_s}$ because it would influence all the measurements in the same direction (adding them would increase the uncertainties of $A_{K_s}/d$ up to 0.04, 0.11, 0.14, and 0.29 mag/kpc). Figure \ref{relative_ext} shows the relation between the extinction and the distance to a given spiral arm. The red dashed lines join the experimental points and correspond to linear fits between consecutive spiral arms, whose slope is the differential extinction per kiloparsec computed previously.

We find an increase in the differential extinction per kiloparsec along the line of sight towards the GC. We also find that the increase in ratio ($r_{i,j}= (A_{K_s}/d)_i/(A_{K_s}/d)_j$) between consecutive arms ($i$,$j$) is $r_{1,2} = 4.00\pm0.07$, $r_{2,3} = 1.54\pm0.24$, and $r_{3,4} = 1.04\pm0.58$. This might indicate an increase in material between arms that would be in agreement with the exponential thin disc of the Milky Way \citep[e.g.][]{Bland-Hawthorn:2016aa}. Other possibilities, such as different grain composition of the interstellar dust, and/or different spiral arm width for the observed line of sight, might also be possible. A number of recent studies aim at analysing the dust distribution along the Galactic plane  \citep[e.g.][]{Rezaei-Kh.:2018aa,Green:2019aa,Lallement:2019aa,Hottier:2020aa}. These are normally limited by a 3 kpc sphere imposed by the limitations of the second release of the Gaia survey. Nevertheless, we can compare their findings with our results for the closest spiral arms. Although the dust distribution seems to be very patchy and does not obviously follow continuous spiral arm footprints, it is possible to see an increase in extinction in the line of sight towards the GC (see Fig. 7 in \citealt{Hottier:2020aa}) and an increase in dust density (Fig. 4 in \citealt{Rezaei-Kh.:2018aa}), which might be in agreement with our results. Deeper photometry and distance measurements are needed for a more sensitive analysis of the innermost regions of the Galactic disc. Our analysis and the GALACTICNUCLEUS survey might be useful for the next steps in tracing and measuring the dust distribution and extinction along the line of sight towards the GC.

\section{Stellar population analysis}
\label{KLF}

\subsection{Luminosity function}

We created de-reddened $K_s$-luminosity functions (KLFs) and fitted them with a linear combination of theoretical luminosity functions (Parsec models) to determine the stellar population and the SFH of the detected spiral arm features. The analysis of the KLFs is based on the relative weight of the main features appearing on the luminosity functions, such as the RC, the red giant branch bump, or the asymptotic giant branch bump \citep[e.g.][]{Nogueras-Lara:2019ad,Schodel:2020aa}. Given that the uncertainties of the luminosity functions are proportionally related to the detected number of stars ($\sim \#stars^{1/2}$), a sufficient number of stars is needed to apply the proposed methodology. Therefore, we excluded the first spiral arm from the analysis. Moreover, some contamination from the Galactic bulge might affect the fourth spiral arm given the proximity in extinction and the use of the CMD $K_s$ versus $H-K_s$, to select the stars belonging to each spiral arm. Hence, we restricted the analysis to the second and the third spiral arms.

First of all, to correct possible saturation problems and also include stars with $K_s<11.5$\,mag, we used the SIRIUS IRSF catalogue \citep[e.g.][]{Nagayama:2003fk,Nishiyama:2006tx} to replace the $K_s$ photometry of stars with $K_s<11.5$\,mag. We accounted for possible deviations of the ZP, correcting the photometry from SIRIUS with respect to non-saturated common bright stars between both catalogues. We also included the photometry of bright stars that were not detected in $K_s$ band in the GALACTICNUCLEUS catalogue because of saturation. We selected stars within the boxes used to create the extinction maps and also included the stars belonging to the ascending giant branch following the isochrones, as indicated by the blue dashed lines in Fig. \ref{ext_maps_selection}. We established a lower limit of $K_s=9$\,mag to account for the saturation of the SIRIUS survey \citep{Matsunaga:2009qp}. In this way, we are able to include more red giant stars whose main features in the KLF, in particular the RC feature \citep[e.g.][]{Girardi:2016fk}, are very useful for disentangling the different stellar populations \citep[e.g.][]{Nogueras-Lara:2018ab,Nogueras-Lara:2019ad}.

We corrected the reddening, applying the extinction maps derived previously for the corresponding spiral arms (Sect.\,\ref{extinction_m}). The KLFs were computed selecting the bin-width using the Python function numpy.histogram \citep{Harris:2020aa}. We used the completeness solution due to crowding obtained in \citet{Nogueras-Lara:2020aa} for the $K_s$ band. We limited the correction to magnitudes where the completeness is $>70\%$. 

To fit the KLFs, we used a similar approach as in \citet{Nogueras-Lara:2019ad,Schodel:2020aa}. We generated 1000 MC samples from the original KLFs, assuming Gaussian uncertainties and fit them with a linear combination of 14 theoretical PARSEC luminosity functions\footnote{http://stev.oapd.inaf.it/cgi-bin/cmd}. We used the following models (assuming metallicities in agreement with \citet{Feltzing:2001aa} for the different ages, as in previous sections, and a Kroupa initial mass function) to sample the age space: 12, 9, 7, 5, 3, 2, 1, 0.7, 0.5, 0.3, 0.1, 0.06, 0.03, and 0.01\,Gyr. We considered a free parameter in the fit for a Gaussian smoothing to take into account different distances to the stars and/or possible photometric uncertainties. We also fitted the distance modulus allowing it to vary in a 3\,$\sigma$ range from the values measured in Table \ref{model_results} for the four spiral arms $J-H$ case. 

\begin{figure}
   \includegraphics[width=\linewidth]{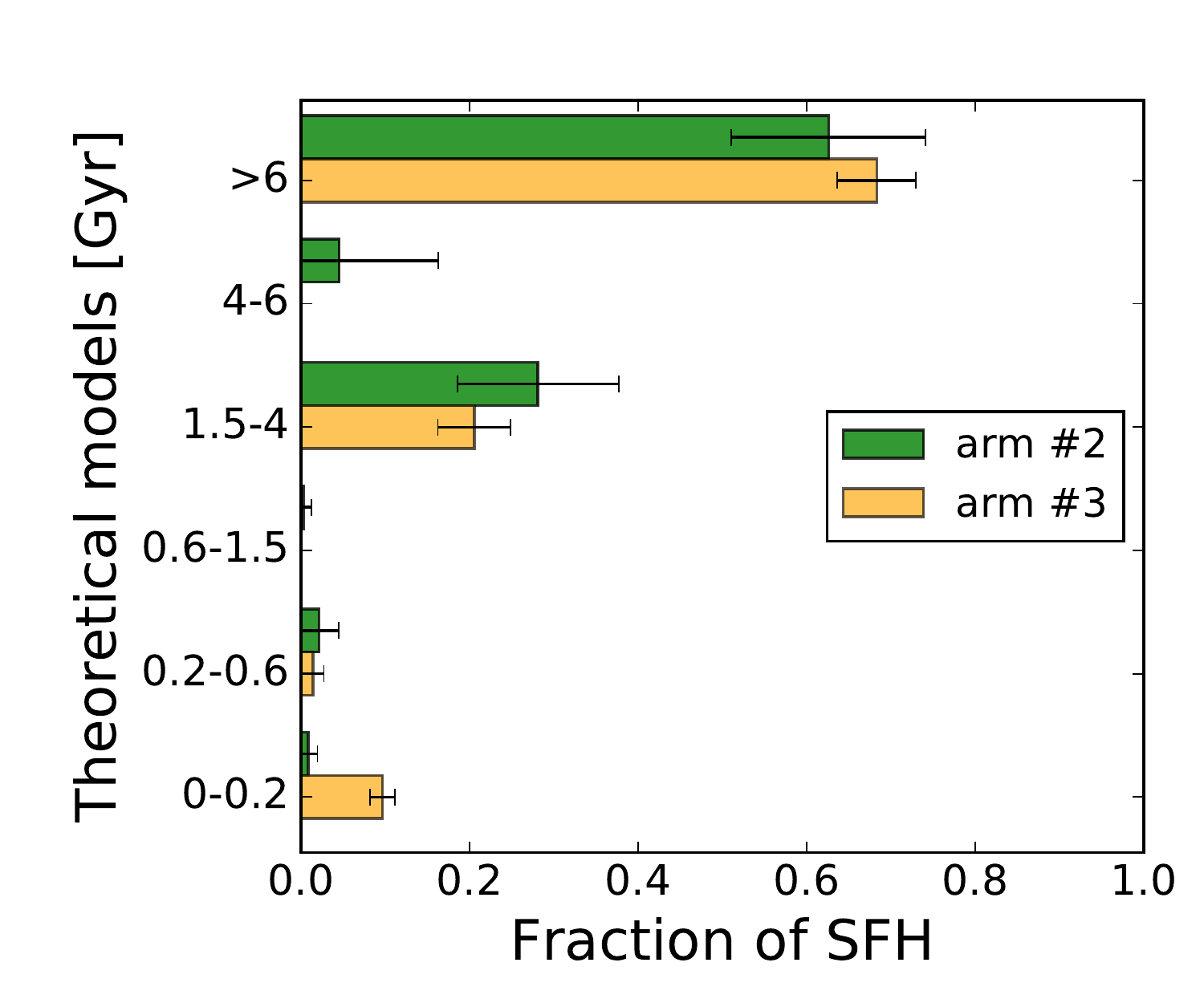}
   \caption{SFHs derived for the second and the third spiral arms, as derived from the MC simulations. The error bars show the 1\,$\sigma$ uncertainties. The absence of any bar and the error bar indicate that there is no contribution for a given age bin.}  
   
       \label{KLF_2_3_res}
    \end{figure}

\begin{figure}
   \includegraphics[width=\linewidth]{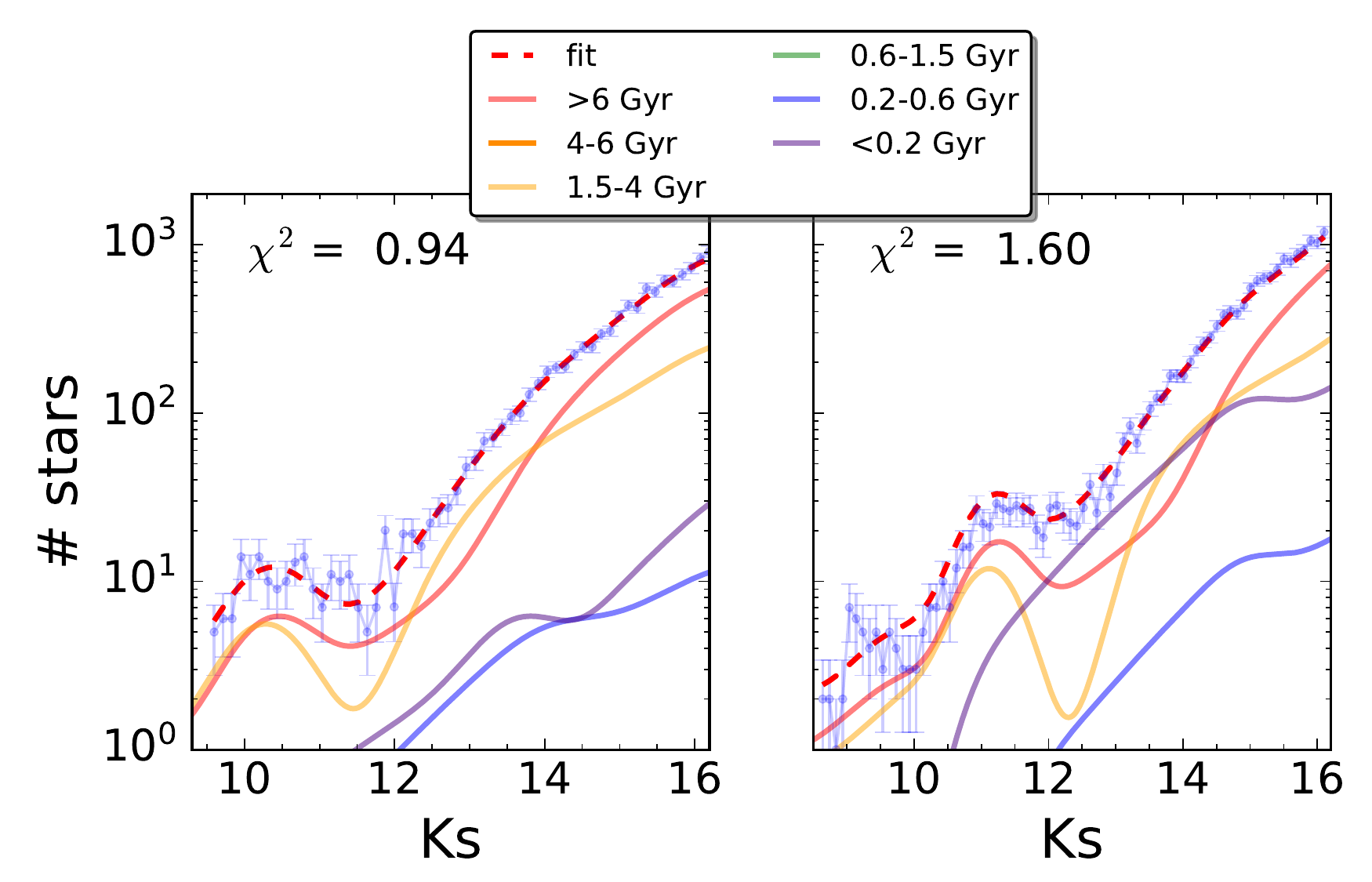}
   \caption{Best-fit solutions obtained for the original (no MC samples) of the KLFs for the second and third spiral arms (left and right panels, respectively).The legend indicates the colour coding corresponding to the selected age bins. The $\chi^2_{red}$ is indicated for each of the cases.}  
   
       \label{KLF_2_3}
    \end{figure}

The best fit for each of the MC realisations was obtained via $\chi^2$ minimisation. To avoid the degeneracies between similar ages (that is more important for old stellar populations), we created larger bins for the ages adding the contribution from several models. Figure \ref{KLF_2_3_res} shows the selected bins and the results obtained for the second and third spiral arms when averaging over the 1000 MC realisations, removing 3\,$\sigma$ outliers. The uncertainties were computed as the standard deviation of the underlying distributions. Figure \ref{KLF_2_3} shows the best fit obtained for the original KLFs, in agreement with the results of the MC simulations.

\subsection{Systematic uncertainty}

To check our results, we explored potential sources of systematic uncertainty. In particular, we varied the bin width of the original KLFs considering a variation of $\pm25\%$. We repeated the MC simulations without obtaining any significant variation within the uncertainties. We also analysed the influence of the models in our results. For this, we used the updated BaSTI models\footnote{http://basti-iac.oa-abruzzo.inaf.it/isocs.html} \citep{Pietrinferni:2004aa,Pietrinferni:2013aa,Bedin:2005aa,Cordier:2007aa,Hidalgo:2018aa} with solar metallicity, a Salpeter initial mass function, and slightly different ages (13, 11, 9, 7, 5, 3, 1, 0.8, 0.5, 0.25, 0.15, 0.1, 0.08, and 0.03\,Gyr). We obtained similar results within the uncertainties for all the stellar ages, except for the second spiral arm, where the bin for ages between 0.6 and 1.5\,Gyr appears to have a somewhat larger ($\sim 5\pm3\%$) contribution. This is not surprising given the lower $K_s$ range used to derive the SFH. In particular, the KLF does not cover the asymptotic giant branch bump and the RC feature is not fully included, which might introduce some bias when deriving the SFH. On the other hand, these features are present in the KLF belonging to the third spiral arm ($\sim9.5$ and $\sim11.5$ mag, respectively).

\subsection{Results}

We found relatively similar stellar  populations for both spiral arms. We identified two main episodes of star formation: (1) the majority of the stellar mass ($\sim60-70\,\%$) was formed more than 6\,Gyr ago. (2) A second important burst appears for ages between 1.5 and 4\,Gyr, where $\sim 20-30\%$ of the stellar mass originated. We also measured some significant star formation at more recent ages for the case of the third spiral arm, accounting for $\sim 10\%$ of the stellar mass. 

Our results agree well with previous stellar population studies on the Milky Way disc. In this sense, \citet{Rocha-Pinto:2000aa} detected three enhanced episodes of star formation of the Milky Way disc around 0-1\,Gyr, 2-5\,Gyr, and 7-9\,Gyr ago that agree well with our findings, in particular with the SFH for the third spiral arm, where we also detected clear signatures of recent star formation. On the other hand, the more recent work by \citet{Ruiz-Lara:2020aa} analysed the stellar population within a radius of $\sim 2$\,kpc from the Sun and found three significant star formation events around 5.7, 1.9, and 1.0\,Gyr ago. The first two agree with the SFH derived in this work, whereas the third one (which is lower in their analysis) appears to be shifted towards younger ages in our results, and/or not to be very important for the case of the second spiral arm. Overall, our results point towards a similar SFH in the solar neighbourhood to that in the spiral arms towards the GC that we analyse here. This also indicates that the models used to fit the spiral arms in Sect. \ref{cmd_fitting} are adequate.

\section{Conclusions}

In the present study, we computed the distance to the spiral arms along the line of sight towards the GC and analysed the extinction in the NIR using the data from the GALACTICNUCLEUS survey \citep{Nogueras-Lara:2018aa, Nogueras-Lara:2019aa}. We investigated the spiral arm structure by fitting a synthetic model to the CMD $H$ versus $J-H$. We applied a $\chi^2$ minimisation technique and computed the distances and extinction to the spiral-arm features, obtaining: $d_1 = 1.6\pm 0.2$ kpc, $d_2 = 2.6\pm 0.2$ kpc, $d_3 = 3.9\pm0.3$ kpc, and $d_4 = 4.5 \pm 0.2$ kpc, and $A_{H1} = 0.35\pm0.08$ mag, $A_{H2} = 0.77\pm0.08$ mag, $A_{H3} = 1.68\pm0.08$ mag, and $A_{H4} = 2.30\pm0.08$ mag. The sub-indices refer to the corresponding spiral arms. 

We checked that our results are robust and addressed potential sources of systematic uncertainty, repeating the calculation using different ages for the synthetic models, bin sizes, photometric bands, SFHs, and simulated models. We also considered the possible effect of the ZP systematic error. Moreover, we analysed the CCD distribution of the population in front of the GC corresponding to the central region of the GALACTICNUCLEUS survey and find that it is compatible with the four-spiral-arms structure.  Finally, we cross-correlated the GALACTICNUCLEUS data with the Gaia DR2 catalogue \citep{Gaia-Collaboration:2018aa} and estimated distances corresponding to the two nearest spiral arms. These are consistent with our results.

We calculated the ratios between the extinction values $A_H$ and $A_{K_s}$, selecting stars belonging to each of the spiral-arm features and using the mean $A_{K_s}$ extinctions derived with the synthetic fitting of the CMD. We conclude that the extinction curve does not significantly change with distance and the mean absolute extinction. Combining the values for the different spiral arms, we obtain $A_J/A_H = 1.91\pm0.11$ and $A_H/A_{K_s} = 1.86\pm0.11$. These values are compatible, within the uncertainties, with the extinction ratios of $A_J/A_{H} = 1.87\pm0.03$ $A_H/A_{K_s} = 1.84\pm0.03$ derived for the GC \citep{Nogueras-Lara:2020aa}, pointing towards a constant extinction law in the studied wavelength regime.

We also created $A_{K_s}$-extinction maps using the features in the CMD $K_s$ versus $H-K_s$ for the central region of the GALACTICNUCLEUS survey. We find that the extinction maps are homogeneous within the uncertainties and do not present any significant variations in the studied line of sight, in contrast to the extinction maps corresponding to the innermost regions of the Galaxy \citep[e.g.][]{Nogueras-Lara:2018aa,Nogueras-Lara:2019ad}. Moreover, we find that the ratio of extinction to distance increases towards the GC, indicating a higher dust density for the inner spiral arms, and/or different dust composition or spiral arm width. We also find that $\sim 50\%$ of the cumulative extinction along the line of sight towards the GC occurs within a distance of $\sim5$ kpc from Earth.

Finally, we studied the stellar populations and the SFH of the second and third spiral arms, analysing their KLFs and fitting them with a linear combination of theoretical models of different ages (Parsec KLFs). We ended up with similar stellar populations for both spiral arms, detecting two main episodes of star formation around $>6$\,Gyr ($\sim 60-70\,\%$ of the stellar mass) and $1.5-4$\,Gyr ($\sim 20-30\,\%$ of the stellar mass), and also more recent star formation ($\sim10\%$) for ages <1\,Gyr for the third spiral arm. Our results agree well with those of previous studies of the Galactic disc stellar population \citep[e.g.][]{Rocha-Pinto:2000aa} and also suggest a similar SFH with respect to the stars within a 2 kpc radius around the Sun \citep{Ruiz-Lara:2020aa}.

  \begin{acknowledgements}
      This work is based on observations made with ESO
      Telescopes at the La Silla Paranal Observatory under programme
      ID 195.B-0283. We thank the staff of
      ESO for their great efforts and helpfulness. 
      
F.N.-L. and N.N. gratefully acknowledge support by the Deutsche Forschungsgemeinschaft (DFG, German Research Foundation) – Project-ID 138713538 – SFB 881 ("The Milky Way System", subproject B8). R.S. acknowledges financial support from the State
Agency for Research of the Spanish MCIU through the "Center of Excellence Severo
Ochoa" award for the Instituto de Astrof\'isica de Andaluc\'ia (SEV-2017-0709). R. S.  acknowledges financial support from national project
PGC2018-095049-B-C21 (MCIU/AEI/FEDER, UE).
\end{acknowledgements}

\bibliographystyle{aa}
\bibliography{../../BibGC}

\appendix
\section{Systematic errors of the synthetic CMD model fitting}
\label{append1}

\begin{table*}
\caption{Influence of bin width on the four-spiral-arm model using the CMD $H$ vs. $J-H$.}
\label{app_bins} 
\begin{center}
\def\arraystretch{1.4}
\setlength{\tabcolsep}{3.8pt}

\begin{tabular}{cc|cc}
 & \multicolumn{1}{c}{} &  & \tabularnewline
\hline 
\hline 
Bin 0.1125-0.225 &  & Bin 0.07-0.14  & \tabularnewline
d (pc) & $A_H$ (mag) & d (pc) & $A_H$ (mag)\tabularnewline
\hline 
1585 $\pm$ 175 & 0.35 $\pm$ 0.08 & 1642 $\pm$ 175 & 0.35 $\pm$ 0.08\tabularnewline
2676 $\pm$ 242 & 0.75 $\pm$ 0.08 & 2743 $\pm$ 284 & 0.76 $\pm$ 0.08\tabularnewline
3836 $\pm$ 288 & 1.68 $\pm$ 0.08 & 3799 $\pm$ 226 & 1.67 $\pm$ 0.08\tabularnewline
4707 $\pm$ 259 & 2.30 $\pm$ 0.08 & 4662 $\pm$ 266 & 2.30 $\pm$ 0.08\tabularnewline
\hline 
$S_{back}$ = 0.86 $\pm$ 0.28 &  & $S_{back}$ = 0.86 $\pm$ 0.28 & \tabularnewline
\hline 
 & \multicolumn{1}{c}{} &  & \tabularnewline
\end{tabular}

\end{center}
\textbf{Notes.} The bin widths indicated in the first row correspond to the $x$- and $y$-axes, respectively. The parameters are analogous to those in Table\,\ref{model_table}.

 \end{table*}

\begin{table*}
\caption{Influence of the ZP systematic errors on the four-spiral-arm model using the CMD $H$ vs. $J-H$.}
\label{app_ZP} 
\begin{center}
\def\arraystretch{1.4}
\setlength{\tabcolsep}{3.8pt}

\begin{tabular}{cc|cc}
 & \multicolumn{1}{c}{} &  & \tabularnewline
\hline 
\hline 
$J+\Delta ZP$  &  & $J-\Delta ZP$ & \tabularnewline
d (pc) & $A_H$ (mag) & d (pc) & $A_H$ (mag)\tabularnewline
\hline 
1544 $\pm$ 175 & 0.46 $\pm$ 0.08 & 1645 $\pm$ 175 & 0.35 $\pm$ 0.08\tabularnewline
2488 $\pm$ 212 & 0.78 $\pm$ 0.08 & 2631 $\pm$ 221 & 0.71 $\pm$ 0.08\tabularnewline
3691 $\pm$ 240 & 1.70 $\pm$ 0.08 & 3688 $\pm$ 294 & 1.62 $\pm$ 0.12\tabularnewline
4538 $\pm$ 188 & 2.30 $\pm$ 0.08 & 4608 $\pm$ 250 & 2.30 $\pm$ 0.08\tabularnewline
\hline 
 $S_{back}$ = 0.77 $\pm$ 0.20 &  &  $S_{back}$ = 0.84 $\pm$ 0.27& \tabularnewline
\hline 
\hline 
$H+\Delta ZP$ &  & $H-\Delta ZP$ & \tabularnewline
d (pc) & $A_H$ (mag) & d (pc) & $A_H$ (mag)\tabularnewline
\hline 
1647 $\pm$ 175 & 0.35 $\pm$ 0.08 & 1536 $\pm$ 175 & 0.47 $\pm$ 0.08\tabularnewline
2654 $\pm$ 223 & 0.71 $\pm$ 0.08 & 2444 $\pm$ 175 & 0.78 $\pm$ 0.08\tabularnewline
3686 $\pm$ 303 & 1.60 $\pm$ 0.11 & 3712 $\pm$ 239 & 1.70 $\pm$ 0.08\tabularnewline
4603 $\pm$ 239 & 2.30 $\pm$ 0.08 & 4507 $\pm$ 175 & 2.30 $\pm$ 0.08\tabularnewline
\hline 
 $S_{back}$ = 0.84 $\pm$ 0.27 &  &  $S_{back}$ = 0.75 $\pm$ 0.16& \tabularnewline
\hline 
\end{tabular}

\end{center}
\textbf{Notes.} $\Delta ZP$ correspond to 0.04 mag in all the cases as indicated in \citet{Nogueras-Lara:2019aa}. The parameters are analogous to those in Table \ref{model_table}.

 \end{table*}

 \begin{table*}
\caption{Influence of the synthetic stellar populations on the four spiral arm model using the CMD $H$ vs. $J-H$.}
\label{app_basti} 
\begin{center}
\def\arraystretch{1.4}
\setlength{\tabcolsep}{3.8pt}

 \begin{tabular}{cc|cc}
 & \multicolumn{1}{c}{} &  & \tabularnewline
\hline 
\hline 
BaSTI & \citet{Ruiz-Lara:2020aa} & BaSTI  & \citet{Rocha-Pinto:2000aa} \tabularnewline
d (pc) & $A_H$ (mag) & d (pc) & $A_H$ (mag)\tabularnewline
\hline 
1650 $\pm$ 318 & 0.29 $\pm$ 0.19 & 1700 $\pm$ 257 & 0.30 $\pm$ 0.17\tabularnewline
2832 $\pm$ 175 & 0.65 $\pm$ 0.08 & 2988 $\pm$ 175 & 0.65 $\pm$ 0.08\tabularnewline
3678 $\pm$ 304 & 1.64 $\pm$ 0.08 & 3756 $\pm$ 288 & 1.66 $\pm$ 0.08\tabularnewline
4601 $\pm$ 213 & 2.28 $\pm$ 0.08 & 4632 $\pm$ 209 & 2.27 $\pm$ 0.08\tabularnewline
\hline 
$S_{back}$ = 0.43 $\pm$ 0.22 &  & $S_{back}$ = 0.67 $\pm$ 0.11 & \tabularnewline
\hline 
 & \multicolumn{1}{c}{} &  & \tabularnewline
\end{tabular}

\end{center}
\textbf{Notes.} The parameters are analogous to those in Table \ref{model_table}.

 \end{table*}

  \begin{table*}
\caption{Simulations to test the reliability of the model fitting.}
\label{app_tests} 
\begin{center}
\def\arraystretch{1.4}
\setlength{\tabcolsep}{3.8pt}

 \begin{tabular}{cccc|cc}
 &  &  & \multicolumn{1}{c}{} &  & \tabularnewline
\hline 
\hline 
Simulated & models &  &  & Fitting & \tabularnewline
d (pc) & $A_H$ & width & scale & d (pc) & $A_H$\tabularnewline
\hline 
1500 & 0.25 & 800 & 1 & 1731 $\pm$ 175 & 0.20 $\pm$ 0.08\tabularnewline
2600 & 0.9 & 700 & 0.8 & 2574 $\pm$ 250 & 0.85 $\pm$ 0.08\tabularnewline
3600 & 1.6 & 860 & 1.1 & 3803 $\pm$ 243 & 1.61 $\pm$ 0.07\tabularnewline
4750 & 2.14 & 1000 & 0.9 & 4720 $\pm$ 300 & 2.19 $\pm$ 0.08\tabularnewline
$S_{back}$ = 0.8  &  &  &  & $S_{back}$ = 1.45 $\pm$ 0.29 & \tabularnewline
\hline 
1700 & 0.18 & 600 & 1 & 1513 $\pm$ 175 & 0.20 $\pm$ 0.08\tabularnewline
2400 & 0.7 & 840 & 1.2 & 2441 $\pm$ 230 & 0.67 $\pm$ 0.08\tabularnewline
3300 & 1.4 & 700 & 0.8 & 3502 $\pm$ 200 & 1.39 $\pm$ 0.04\tabularnewline
4500 & 2 & 840 & 1.1 & 4800 $\pm$ 242 & 2.00 $\pm$ 0.08\tabularnewline
$S_{back}$= 0.9 &  &  &  & $S_{back}$= 0.52 $\pm$ 0.22 & \tabularnewline
\hline 
1400 & 0.33 & 700 & 1 & 1619 $\pm$ 175 & 0.35 $\pm$ 0.08\tabularnewline
2700 & 0.82 & 760 & 0.9 & 2613 $\pm$ 218 & 0.81 $\pm$ 0.08\tabularnewline
3500 & 1.53 & 840 & 1.1 & 3663 $\pm$ 286 & 1.55 $\pm$ 0.08\tabularnewline
4650 & 2.12 & 700 & 0.8  & 4783 $\pm$ 311 & 2.15 $\pm$ 0.08\tabularnewline
$S_{back}$ = 1.1 &  &  &  & $S_{back}$ = 1.21 $\pm$ 0.41 & \tabularnewline
\hline 
 &  &  & \multicolumn{1}{c}{} &  & \tabularnewline
\end{tabular}

\end{center}
\textbf{Notes.} Left columns: Simulated parameters. Right columns: Results obtained with the model fitting. The parameters are analogous to those in Table \ref{model_table}.

 \end{table*}

\end{document}